\documentclass[reprint,amsmath,amssymb,aps,pra]{revtex4-2}
\usepackage{graphicx}
\usepackage{dcolumn}
\usepackage{bm}
\usepackage{xcolor}

\begin{document}
\preprint{APS/123-QED}
\title{Convexity of the longitudinal variation of third-order resonance driving terms and its application in dynamic aperture optimization}
\author{Wanbin Li}
\author{Zihan Wang}%
\author{Yuejing Huang}%
\author{Bingfeng Wei}%
\email{Contact author: weibf@ustc.edu.cn}
\author{Zhenghe Bai}
\email{Contact author: baizhe@ustc.edu.cn}
\affiliation{National Synchrotron Radiation Laboratory, University of Science and Technology of China, Hefei 230029, China}

\date{\today}

\begin{abstract}  
  The optimization of the dynamic aperture (DA) of a storage ring is typically a non-convex problem with multiple local optima. 
  Recent studies showed that reducing the variation of resonance driving terms (RDTs) along the longitudinal position improves DA very effectively, 
  as the reduction in the longitudinal variation of lower-order RDTs suppresses higher-order nonlinear terms. 
  Therefore, minimizing the longitudinal variation of third-order RDTs is crucial for DA optimization. 
  In this paper, we prove that the longitudinal variation of third-order RDTs, 
  quantified using their RMS value $f_{3,\mathrm{rms}}$ at sextupole locations, is a special convex function. 
  In the space of sextupole strengths, 
  the iso-surfaces of $f_{3,\mathrm{rms}}$ are a series of concentric and coaxial ellipsoidal surfaces, 
  with the central position possessing minimum $f_{3,\mathrm{rms}}$. 
  The scanning results of a storage ring lattice show a strong consistency between the distributions of $f_{3,\mathrm{rms}}$ and DA, 
  indicating that the optimization of DA can be regarded as a roughly approximate convex optimization problem. 
  Based on this, a fast DA optimization method based on particle tracking is developed, 
  where a high-quality initial population for an intelligent algorithm is generated with a Gaussian distribution based on the geometric structure of $f_{3,\mathrm{rms}}$. 
\end{abstract}

\maketitle

\section{\label{sec:introduction}Introduction}  
  In the design of a storage ring, 
  dynamic aperture (DA) is a key performance metric that influences beam injection and lifetime. 
  Within the parameter space defined by sextupole strengths, 
  the DA typically exhibits a complex distribution characterized by multiple local optima, 
  representing a typical non-convex optimization problem. 
  Furthermore, as the natural emittance decreases and the number of sextupole families increases, 
  the optimization of DA becomes increasingly challenging.

  The optimization methods for DA can be categorized into numerical and analytical approaches \cite{Borland:jsr2014, Bartolini:ipac2016}. 
  Numerical optimization methods directly optimize DA based on particle tracking using techniques such as the scanning method and intelligent optimization algorithms \cite{Borland:ipac2009, Yang:prab2011, Bai:ipac2011, Ehrlichman:prab2016, Husain:nima2018, Liyongjun:prab2018}.
  The numerical methods have the potential to find the global optimum. 
  However, they require high computational costs, 
  and offer little physical insight to guide further design and optimization. 
  In contrast, analytical optimization methods have the advantage of providing physical guidance. 
  Minimizing resonance driving terms (RDTs) \cite{Bengtsson:report1997, Franchi:prab2014} combined with frequency map analysis (FMA) \cite{Laskar:book1999, Laurent:prab2003} is a widely used analytical approach, 
  in which the DA is enlarged by reducing RDTs and controlling amplitude-dependent tune shifts (ADTSs) \cite{Smith:epac2002, Tian:cpc2009, Cai:prab2012, Bengtsson:article2017, Streun:prab2023}. 
  
  Recently, it was demonstrated that reducing the variation of RDTs along the longitudinal position is much more effective in enlarging DA than reducing the commonly used one-turn RDTs \cite{Wei:prab2023, Wei:prab2024}. 
  The underlying physics was elucidated through a further analysis of the Baker-Campbell-Hausdorff (BCH) formula \cite{Chao:book2022}, 
  which shows that reducing the longitudinal variation of lower-order RDTs effectively reduces higher-order RDTs and controls ADTSs \cite{Wei:prab2023}. 
  Therefore, minimizing the longitudinal variation of third-order RDTs is crucial for DA optimization. 
  
  In this paper, we will further prove that the longitudinal variation of third-order RDTs, 
  quantified using their RMS value at sextupole locations, is a special convex function. 
  Specifically, its iso-surfaces are a series of concentric and coaxial ellipsoidal surfaces in the sextupole strength space.
  Then, it will be shown, using lattice examples, that solutions with better DAs are distributed in the central region of the ellipsoid, 
  even though the optimization of DA is generally a non-convex optimization problem. 
  Building on this convexity, a fast DA optimization method based on particle tracking will be developed.
  
  The remainder of this paper is organized as follows. 
  The theory for RDTs and reducing their longitudinal variation is revisited in Section~\ref{sec:second section},
  followed by a further demonstration using a simple FODO lattice. 
  In Sec.~\ref{sec:third section}, we prove the convexity of the longitudinal variation of third-order RDTs, 
  and discuss the case of higher-order RDTs. 
  Using storage ring lattice examples, 
  Sec.~\ref{sec:fourth section} first demonstrates, via global scanning, 
  the convexity of the third-order RDT variation and compares its distribution with that of DAs; 
  then, a fast numerical method is developed for DA optimization based on this convexity.
  Finally, conclusion and outlook are given in Sec.~\ref{sec:fifth section}.
    
\section{\label{sec:second section}Control of the longitudinal variation of third-order RDTs}
  \subsection{The longitudinal variation of third-order RDTs}
    The one-turn map $\mathcal{M}$ of a storage ring with $N$ thin nonlinear kicks is given by \cite{Bengtsson:report1997} 
    
    \begin{equation}
      \begin{aligned}
        \mathcal{M}_{0\rightarrow N+1} &= \mathcal{A}_{0}^{-1}e^{:\hat{V}_{1}:}e^{:\hat{V}_{2}:}\dotsb e^{:\hat{V}_{N}:}\mathcal{R}_{0\rightarrow N+1}\mathcal{A}_{N+1}\\
          &= \mathcal{A}_{0}^{-1}e^{:h:}\mathcal{R}_{0\rightarrow N+1}\mathcal{A}_{N+1},
      \end{aligned}
      \label{eq:one_turn_map}
    \end{equation}
    where $\mathcal{A}$ is a normalizing map, 
    $\mathcal{R}$ is a phase space rotation, 
    and $e^{:\hat{V}_{n}:}$ and $e^{:h:}$ are the Lie maps of the $n$-th nonlinear kick and the ring, respectively.
    Using the BCH formula, $e^{:h:}$ can be expanded as
    \begin{equation}
      \begin{aligned}
        e^{:h:} &= \exp\bigg( : \sum_{i=1}^{N} \hat{V}_{i} + \frac{1}{2} \sum_{1\leq i<j}^{N} \Big\{\hat{V}_{i}, \hat{V}_{j}\Big\}\\
                &+ \frac{1}{12} \sum_{1\leq i<j}^{N} \bigg[\Big\{\hat{V}_{i}, \{\hat{V}_{i}, \hat{V}_{j}\}\Big\} - \Big\{\hat{V}_{j}, \{\hat{V}_{i}, \hat{V}_{j}\}\Big\} \bigg]\\
                &+ \frac{1}{4}\sum_{k}^{N} \Big\{\sum_{1\leq i<j}^{k-1}\{\hat{V}_{i}, \hat{V}_{j}\}, \hat{V}_{k}\Big\} + \dotsb : \bigg),
      \end{aligned}
      \label{eq:Lie_map}
    \end{equation}
    where $\{\cdot, \cdot\}$ denotes the Poisson bracket. 

    Now we define two quantities to represent the accumulation along the ring of the first two terms in Eq.~\eqref{eq:Lie_map}:
    \begin{align}
      &A_{t}^{(1)} \equiv \sum_{i=1}^{t} \hat{V}_{i},\label{eq:AT_1}\\
      &\begin{aligned}
        A_{t}^{(2)} &\equiv \frac{1}{2}\sum_{1\leq i<j}^{t} \{\hat{V}_{i}, \hat{V}_{j}\} = \frac{1}{2}\sum_{j=2}^{t}\left\{\sum_{i=1}^{j}\hat{V}_{i}, \hat{V}_{j}\right\}\\
            &= \frac{1}{2}\sum_{j=2}^{t}\{A_{j}^{(1)}, \hat{V}_{j}\}.\label{eq:At_2}
      \end{aligned}
    \end{align} 
    It is seen that $A_{t}^{(2)}$ is constructed from $A_{t}^{(1)}$. 
    Combining Eqs.~\eqref{eq:Lie_map}, \eqref{eq:AT_1}, and \eqref{eq:At_2}, 
    it is known that higher-order terms are formed through the accumulation process of lower-order terms.
    Consequently, minimizing the variation of lower-order terms along the ring leads to the reduction of higher-order terms \cite{Wei:prab2023, Wei:ipac2023, Bai:fls2024}. 
    Using the resonance basis, the terms in the expansion of $e^{:h:}$ can be expressed in terms of RDTs \cite{Bengtsson:report1997, Franchi:prab2014}. 
    Correspondingly, minimizing the longitudinal variation of third-order RDTs can effectively reduce higher-order RDTs and thus enlarge the DA. 
    
    In this paper, we only study the on-momentum DA. 
    The geometric RDT $f_{jklm}(z)$ at an observation position $z$ between the $n$-th and $(n+1)$-th sextupoles is given by \cite{Franchi:prab2014}
    \begin{align}
      \begin{aligned}
        f_{jklm}(z) = \frac{\sum_{w=n+1}^{n+N} h_{w,jklm} e^{i[(j-k)\Delta\phi_{w,x}^{(z)} + (l-m)\Delta\phi_{w,y}^{(z)}]}}{1 - e^{2\pi i[(j-k)\nu_{x} + (l-m)\nu_{y}]}},
        \label{eq:f_jklm}
      \end{aligned}
    \end{align}
    where $h_{w,jklm}$ is a coefficient dependent on the strength of the $w$-th sextupole and the beta functions at the sextupole location,  
    $\Delta\phi_{w,x}^{(z)}$ and $\Delta\phi_{w,y}^{(z)}$ are the horizontal and vertical phase advances between the $w$-th sextupole and the observation position $z$, 
    and $\nu_{x}$ and $\nu_{y}$ are the horizontal and vertical tunes, respectively.     
    In this paper, the longitudinal variation of RDTs is quantified using the RMS value of their magnitudes at sextupole locations, as in Ref.~\cite{Wei:phd2025}. 
    For the longitudinal variation of third-order RDTs, it is calculated as:
    \begin{equation}
      \begin{aligned}
        &f_{3,\mathrm{rms}} = \sqrt{\sum_{3=j+k+l+m}f_{jklm,\mathrm{rms}}^{2}},\\
        \text{with}\quad &f_{jklm,\mathrm{rms}} = \sqrt{\frac{1}{N}\sum_{i=1}^{N}|f_{jklm}(z_{i})|^{2}},
      \end{aligned}
      \label{eq:f3_rms}
    \end{equation}
    where $f_{jklm,\mathrm{rms}}$ is the RMS for a single third-order RDT, 
    and $z_{i}$ is the position of the $i$-th sextupole.
  
  \subsection{\label{sec:FODO}Example: a FODO lattice}
\begin{figure}[htbp]
      \includegraphics[width=0.45\textwidth]{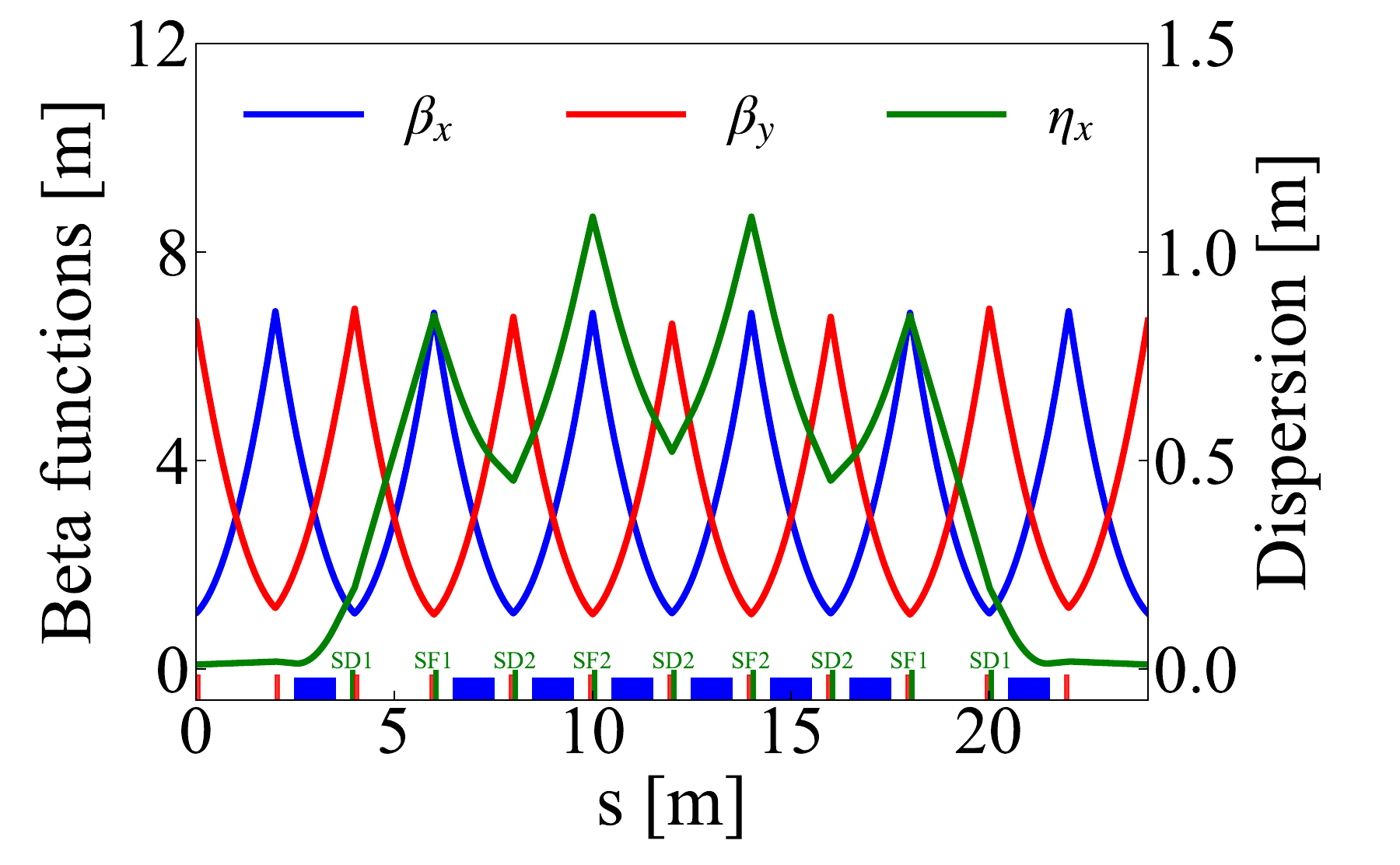}
      \caption{
            Magnet layout and optical functions of a FODO lattice cell, one quarter of the simply designed ring with thin-lens quadrupoles and sextupoles. 
            The four families of chromatic sextupoles indicated in the layout are placed at the same locations as quadrupoles.
        }
      \label{fig:fodo_lattice}
    \end{figure}

    A simply designed FODO lattice, shown in Fig.~\ref{fig:fodo_lattice},
    was used to illustrate the relationship between reducing the longitudinal variation of RDTs and improving the DA.
    In this lattice, both the quadrupoles and sextupoles employ a thin-lens model. 
    We randomly generated 500 nonlinear solutions for study, 
    with the horizontal and vertical chromaticities corrected to (1.0, 1.0). 
    
    For each nonlinear solution, the DA was tracked through FMA using the \texttt{elegant} code \cite{Borland:icap2000}, 
    obtaining the number of surviving particles and their frequency diffusion rates.
    The number of surviving particles, denoted as $N_{\mathrm{particle}}$, represents the size of DA.
    The frequency diffusion rate, indicating the stability of particle motion, is computed as
    \begin{align}
      d_{r} = \log_{10}\Big(\sqrt{\Delta\nu_{x}^{2} + \Delta\nu_{y}^{2}}\Big/N_{\text{turn}}\Big),
    \end{align}
    where $N_{\text{turn}}$ is the total number of tracking turns, 
    and $\Delta\nu_{x}$ and $\Delta\nu_{y}$ are the frequency differences in the $x$ and $y$ planes between the first and second half of tracking turns, respectively. 
    To represent the quality of DA, we define $\overline{d_{r}}$, which is the average frequency diffusion rate of all surviving particles. 
    Therefore, $N_{\mathrm{particle}}$, together with $\overline{d_{r}}$, provides a comprehensive characterization of the DA.
    
\begin{figure}[htbp]
        \centering
        \includegraphics[width=0.45\textwidth]{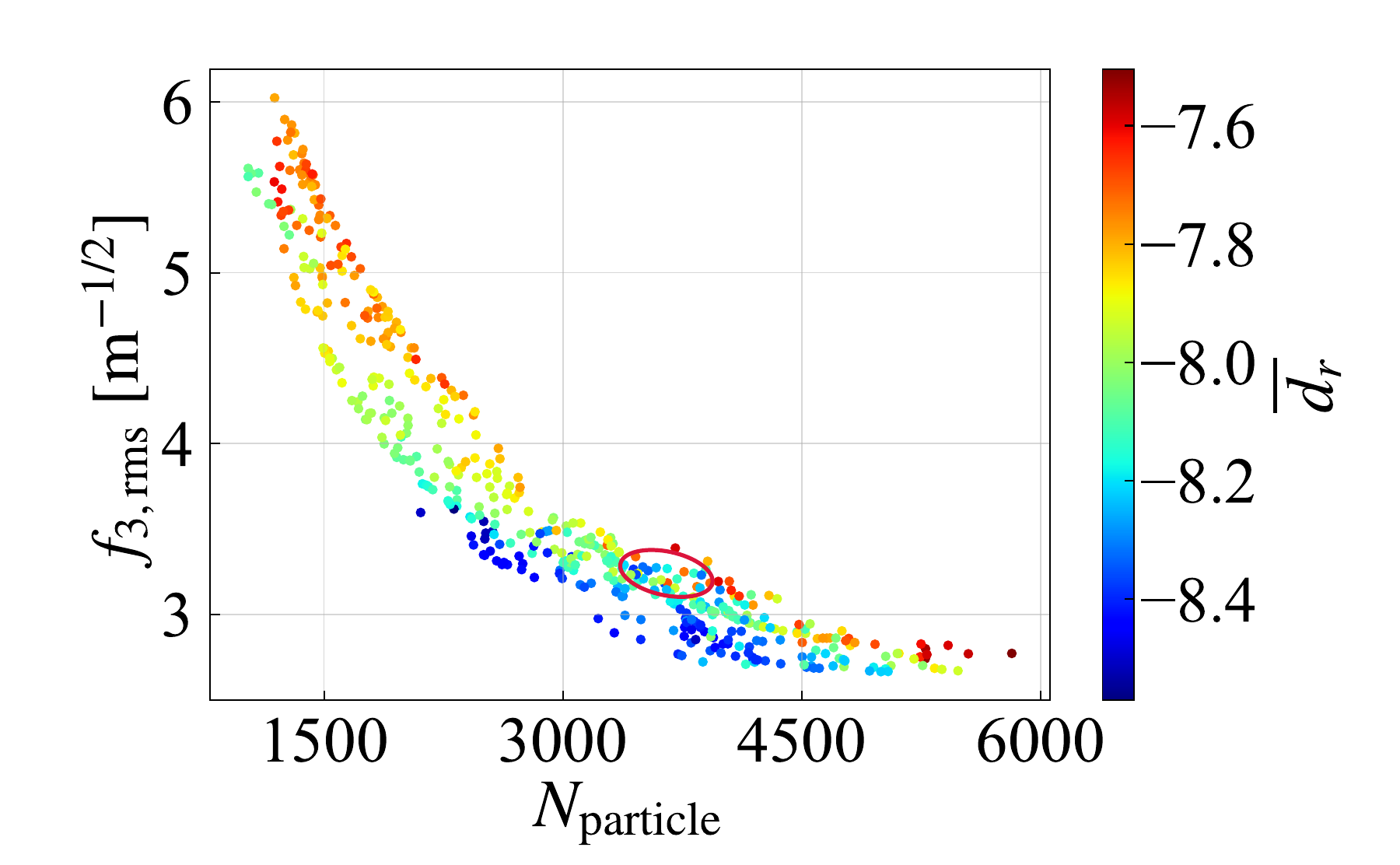}
        \caption{
            Relationship between $f_{3,\mathrm{rms}}$, $N_{\text{particle}}$ and $\overline{d_{r}}$ for the nonlinear solutions of the FODO lattice.
            The blue solutions within the red ellipse are influenced by the ADTS terms.
        }
        \label{fig:fodo_DA_f3_dr}
    \end{figure}

    Figure~\ref{fig:fodo_DA_f3_dr} illustrates the relationship between $f_{3,\mathrm{rms}}$, 
    $N_{\text{particle}}$, and $\overline{d_{r}}$ for the nonlinear solutions. 
    It is clearly seen that (1) $N_{\text{particle}}$ increases with the reduction of $f_{3,\mathrm{rms}}$, 
    and that (2) for a given $N_{\text{particle}}$, $\overline{d_{r}}$ decreases with the reduction of $f_{3,\mathrm{rms}}$.
    This indicates that reducing $f_{3,\mathrm{rms}}$ enlarges the DA and also enhances the stability of particle motion, 
    thereby improving the overall DA performance. 
    In the figure, the blue solutions within the red ellipse are an exception due to smaller ADTSs.

    \begin{figure}[htbp]
      \includegraphics[width=0.45\textwidth]{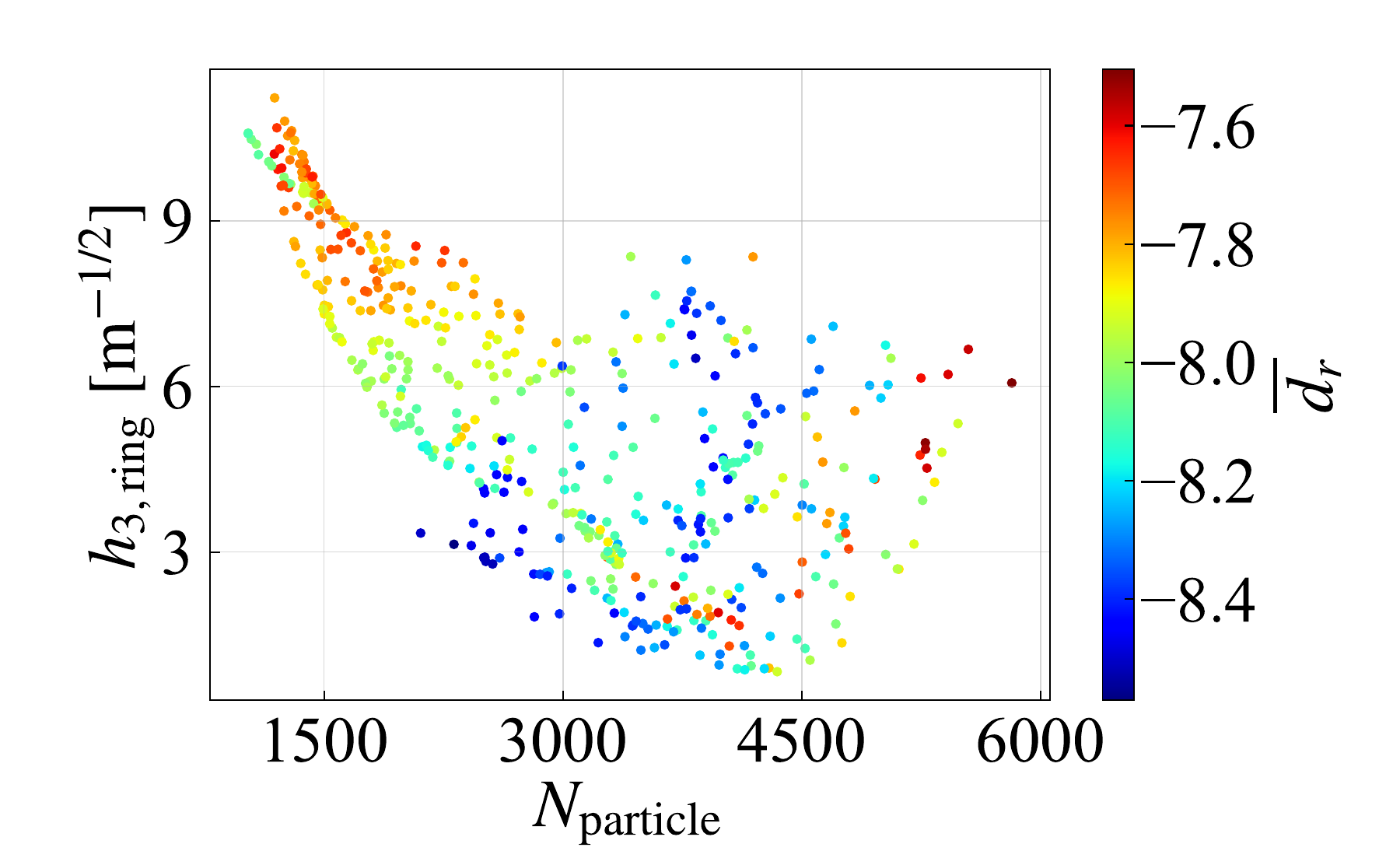}
      \caption{Relationship between $h_{3,\text{ring}}$, $N_{\text{particle}}$ and $\overline{d_{r}}$ for the nonlinear solutions of the FODO lattice.}
      \label{fig:fodo_DA_h3_dr}
    \end{figure}

    The relationship between the third-order one-turn RDTs, 
    $N_{\text{particle}}$, and $\overline{d_{r}}$ for the nonlinear solutions is presented in Fig.~\ref{fig:fodo_DA_h3_dr}.     
    In the figure, the third-order one-turn RDTs are quantified by their RMS value, denoted as $h_{3,\text{ring}}$ \cite{Wei:prab2023}. 
    It is seen that (1) the relationship between $h_{3,\text{ring}}$ and $N_{\text{particle}}$ is significantly weaker than that between $f_{3,\mathrm{rms}}$ and $N_{\text{particle}}$, 
    and that (2) there is no clear relationship between $h_{3,\text{ring}}$ and $\overline{d_{r}}$. 
    Therefore, comparing Figs.~\ref{fig:fodo_DA_f3_dr} and \ref{fig:fodo_DA_h3_dr} clearly shows that reducing the longitudinal variation of third-order RDTs is significantly more effective in improving the DA performance than reducing the commonly used one-turn RDTs. 
    
\section{\label{sec:third section}Convexity of the longitudinal variation of third-order RDTs}
  As is well known, the DA optimization is generally a non-convex optimization problem. 
  However, as will be proven, $f_{3,\mathrm{rms}}$, quantifying the longitudinal variation of third‑order RDTs, 
  is a special convex function. 

  \subsection{\label{sec:defination}Definition of the convex function}
    An optimization problem can be formulated in the form \cite{Boyd:book2004}:
    \begin{equation}
      \begin{aligned}
        \text{minimize}\quad &f_{0}(\vec{x}),\\
        \text{subject to}\quad &f_{i}(\vec{x}) \leq b_{i}, \quad i = 1, \cdots, m.
        \label{eq:optimization_problem}
      \end{aligned}
    \end{equation}
    Here the vector $\vec{x}=(x_{1},\cdots,x_{n})$ is the optimization variable; 
    the functions $f_{0}(\vec{x})$ and $f_{i}(\vec{x})$ are the objective and constraint functions, respectively; 
    and the constants $b_{1},\cdots,b_{m}$ are the bounds for the constraints.
    A convex optimization problem is one in which the objective and constraint functions are convex, 
    which means that they satisfy
    \begin{align}
      f_{i}(\alpha \vec{x} + \beta \vec{y}) \leq \alpha f_{i}(\vec{x}) + \beta f_{i}(\vec{y})
      \label{eq:convex_inequality}
    \end{align}
    for all $\vec{x},\ \vec{y} \in \mathbf{R}^{n}$ and all $\alpha,\ \beta \in \mathbf{R}$ with $\alpha + \beta = 1$, 
    $\alpha\geq 0, \ \beta\geq 0$ \cite{Boyd:book2004}.

    \begin{figure}[htbp]
      \includegraphics[width=0.45\textwidth]{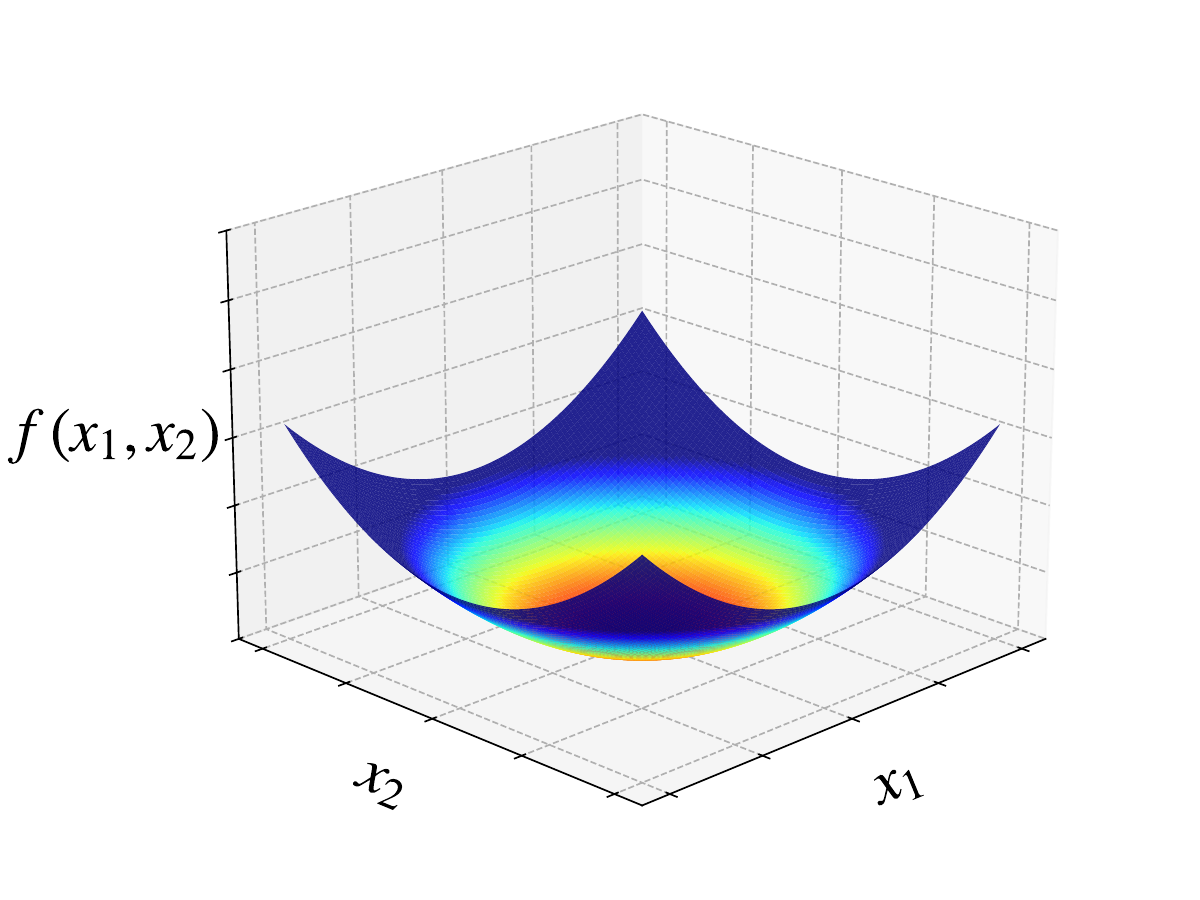}
      \caption{A three-dimensional plot depicting a simple convex quadratic function $f(x_{1},x_{2})=x_{1}^{2}+x_{2}^{2}$.}
      \label{fig:convex_function}
    \end{figure}
    
    A function $f$ is convex if and only if its domain is a convex set and its Hessian $H$ or second derivative $\nabla^{2} f$ is positive semi-definite, expressed as
    \begin{align}
      H = \nabla^{2} f(\vec{x}) \succeq 0
      \label{eq:second_order_condition}
    \end{align}
    for all $\vec{x}$ in the domain \cite{Boyd:book2004}. 
    As an example, Fig.~\ref{fig:convex_function} shows a strictly convex quadratic function. 
    Its Hessian matrix is positive-definite (i.e., $H\succ 0$), 
    which means that it has a unique minimum point within its domain. 

  \subsection{\label{sec:proof}Proof of the convexity of $f_{3,\mathrm{rms}}$}     
    Now we will first prove that, 
    without considering the constraint of chromaticity correction, 
    $f_{3,\mathrm{rms}}$ is a special convex function defined on the $N$-dimensional real space of sextupole strengths, 
    characterized by ellipsoidal iso-surfaces in the sextupole strength space. 
    Then, when the constraint of chromaticity correction is considered, 
    $f_{3,\mathrm{rms}}$ preserves convexity on its feasible set,
    but its iso-surfaces become ($N\!-\!2$)-dimensional ellipsoidal surfaces.
    
    For a third-order geometric RDT generated by normal sextupoles, the coefficient $h_{w,jklm}$ of $f_{jklm}(z)$ in Eq.~\eqref{eq:f_jklm} is given by \cite{Franchi:prab2014}
    \begin{align}
      \begin{aligned}
        h_{w,jklm} &= -\frac{i^{l+m}K_{w}(\beta_{w,x})^{\frac{j+k}{2}}(\beta_{w,y})^{\frac{l+m}{2}}}{8\,\ j!\,\ k!\,\ l!\,\ m!} ,
      \end{aligned}
      \label{eq:h_jklm}
    \end{align}
    where $K_{w}$ is the normalized integrated strength of the $w$-th sextupole,
    and $\beta_{w,x}$ and $\beta_{w,y}$ are respectively the horizontal and vertical beta functions at the sextupole location.     
    Combining Eqs.~\eqref{eq:f_jklm} and \eqref{eq:h_jklm}, 
    it is known that $f_{jklm}(z)$ is a linear combination, 
    with complex coefficients, of sextupole strengths.
    Then, $|f_{jklm}(z)|^{2}$ can be simplified into the following form:
    \begin{equation}
      \begin{aligned}
        |f_{jklm}(z)|^{2} &= \left[ \Re\, f_{jklm}(z) \right]^{2} + \left[ \Im\, f_{jklm}(z) \right]^{2}\\
                        &= (A_{1}K_{1} + \dotsb + A_{n}K_{n} + \dotsb + A_{N}K_{N})^{2}\\
                        &\quad+ (B_{1}K_{1} + \dotsb + B_{n}K_{n} + \dotsb + B_{N}K_{N})^{2}\\
                        &= C_{11}K_{1}^{2} + \dotsb + C_{ij}K_{i}K_{j} + \dotsb + C_{NN}K_{N}^{2}, 
      \end{aligned}
      \label{eq:f_jklm_magnitude_2}
    \end{equation}
    where $A_{n}$, $B_{n}$ and $C_{ij}$ are real coefficients, and $C_{ij}=C_{ji}$. 
    Further combining Eqs.~\eqref{eq:f3_rms} and \eqref{eq:f_jklm_magnitude_2}, $f_{3,\mathrm{rms}}^{2}$ has the following form:
    \begin{equation}
      \begin{aligned}
        &f_{3,\mathrm{rms}}^{2} =\sum_{3=j+k+l+m}\left(\frac{1}{N}\sum_{i=1}^{N}|f_{jklm}(z_{i})|^{2}\right)\\
                    &= D_{11}K_{1}^{2} + \dotsb + D_{ij}K_{i}K_{j} + \dotsb + D_{NN}K_{N}^{2}\\
                    &= \begin{pmatrix}K_{1}, K_{2}, \cdots, K_{N}\end{pmatrix}
                        \begin{pmatrix}
                            D_{11} & D_{12} & \cdots & D_{1N}\\
                            D_{21} & D_{22} & \cdots & D_{2N}\\
                            \vdots & \vdots & \ddots & \vdots\\
                            D_{N1} & D_{N2} & \cdots & D_{NN}\\
                        \end{pmatrix}
                        \begin{pmatrix}K_{1}\\ K_{2}\\ \vdots\\ K_{N}\\\end{pmatrix}\\
                    &= \mathbf{K}^{\mathrm{T}}\mathbf{DK},
      \end{aligned}
      \label{eq:f3_rms_2}
    \end{equation}
    where $\mathbf{K}$ is an $N$-dimensional real vector of sextupole strengths without considering the constraint of chromaticity correction, 
    and \(\mathbf{D}\) is a square matrix whose elements \(D_{ij}\) are real coefficients determined by the lattice functions. 

\begin{figure}[htbp]
      \includegraphics[width=0.45\textwidth]{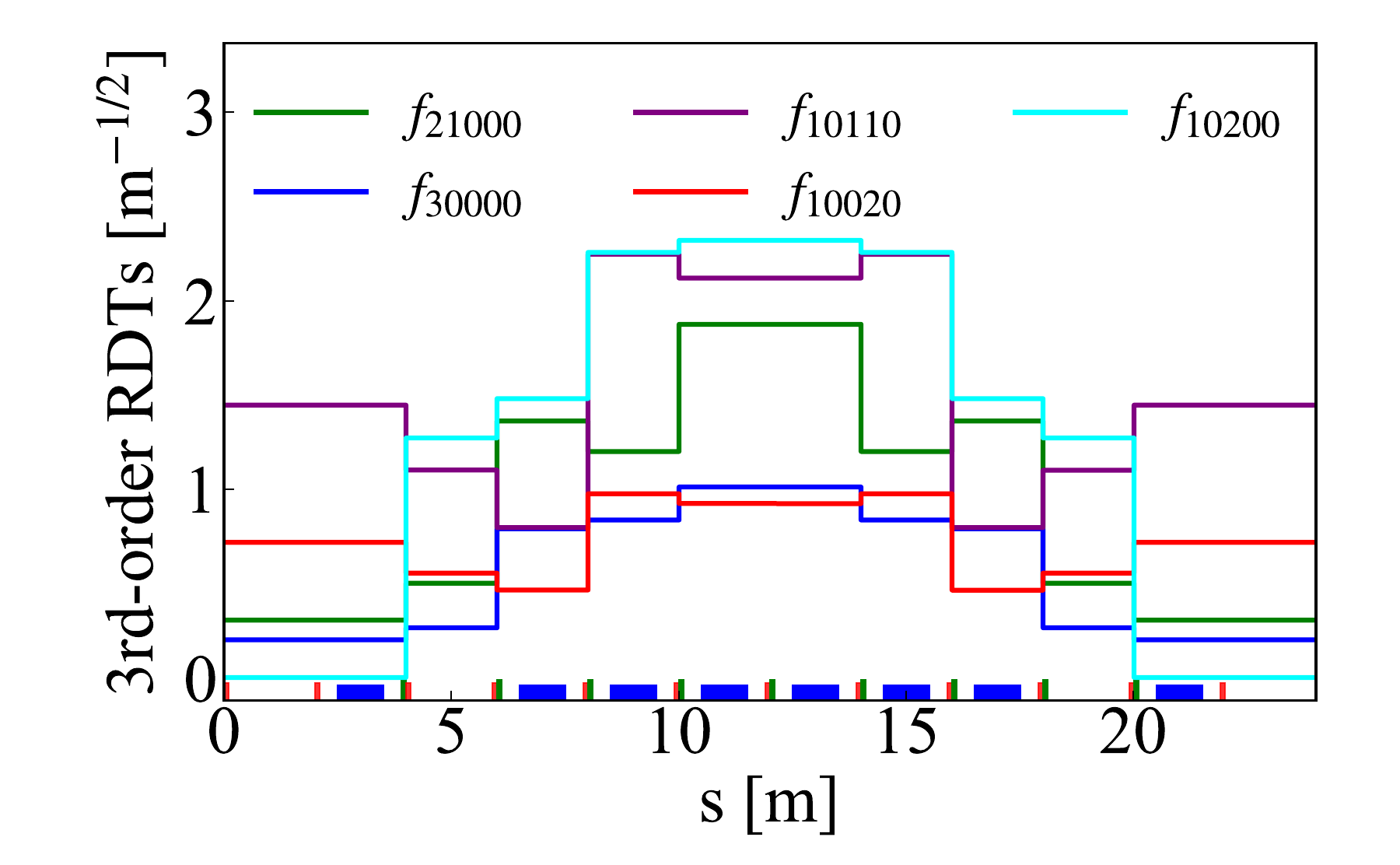}
      \caption{Magnitude variation of third-order RDTs along the longitudinal position for a nonlinear solution of the FODO lattice.}
      \label{fig:fodo_f_jklm}
    \end{figure}

    Since $C_{ij} = C_{ji}$ in Eq.~\eqref{eq:f_jklm_magnitude_2}, it follows that $D_{ij} = D_{ji}$ in Eq.~\eqref{eq:f3_rms_2}, 
    which means that $\mathbf{D}$ is a real symmetric matrix. 
    Since $|f_{jklm}(z)|$ is nonnegative, 
    as illustrated in Fig.~\ref{fig:fodo_f_jklm} using a nonlinear solution of the FODO lattice as an example, 
    the quadratic form $f_{3,\mathrm{rms}}^{2} = \mathbf{K}^{\mathrm{T}}\mathbf{DK}$ is positive for all nonzero $\mathbf{K}$.
    Hence, $\mathbf{D}$ is a positive definite matrix \cite{Horn:book2012}. 
    Consequently, $f_{3,\mathrm{rms}}$ can be written as
    \begin{align}
      f_{3,\mathrm{rms}} = \sqrt{\mathbf{K}^{\mathrm{T}}\mathbf{DK}} = \|\mathbf{K}\|_{\mathbf{D}},
      \label{eq:f3_rms_norm}
    \end{align}
    where $\|\mathbf{K}\|_{\mathbf{D}}$ is a quadratic norm \cite{Boyd:book2004}.
    Every norm on an $N$-dimensional real space $\mathbb{R}^{N}$ is convex, 
    and the unit ball of a quadratic norm is an ellipsoid \cite{Boyd:book2004}. 
    Therefore, $f_{3,\mathrm{rms}}$ is a convex function defined on $\mathbb{R}^{N}$, 
    with its iso-surfaces forming a series of concentric and coaxial ellipsoidal surfaces in the variable space.
    
    In the lattice design, chromaticities are generally required to be corrected to small positive values. 
    This requirement imposes constraints on the strengths of sextupoles, 
    which satisfy the chromaticity correction equations in the horizontal and vertical planes: 
    \begin{equation}
    \left\{
      \begin{aligned}
        \Delta\xi_{x} &= \xi_{x}^{(1)} - \xi_{x}^{(0)} = \frac{1}{4\pi}\sum_{i=1}^{N} \beta_{i,x}\eta_{i,x}K_{i}\\
            &= (E_{1}, \dotsb, E_{i}, \dotsb, E_{N})\mathbf{K} = \mathbf{E}^{\mathrm{T}}\mathbf{K},\\
        \Delta\xi_{y} &= \xi_{y}^{(1)} - \xi_{y}^{(0)} = -\frac{1}{4\pi}\sum_{i=1}^{N} \beta_{i,y}\eta_{i,x}K_{i}\\
            &= (F_{1}, \dotsb, F_{i}, \dotsb, F_{N})\mathbf{K} = \mathbf{F}^{\mathrm{T}}\mathbf{K}.\\
      \end{aligned}
    \right.
    \label{eq:chromaticity_correction}
    \end{equation}
    Here, $\xi_{x}^{(0)}$ and $\xi_{y}^{(0)}$ are the horizontal and vertical natural chromaticities, respectively,  
    $\xi_{x}^{(1)}$ and $\xi_{y}^{(1)}$ are the corrected chromaticities, 
    and $E_{i}$ and $F_{i}$ are real coefficients determined by the dispersion $\eta_{i,x}$ and beta functions at the location of the $i$-th sextupole. 
    Under the two linear equality constraints above, the feasible set of $f_{3,\mathrm{rms}}$ is 
    \begin{align}
        \{\mathbf{K}\in\mathbb{R}^{N}\mid\Delta\xi_{x} = \mathbf{E}^{\mathrm{T}}\mathbf{K},\ \Delta\xi_{y} = \mathbf{F}^{\mathrm{T}}\mathbf{K}\},\nonumber
        \label{eq:feasible_set}
    \end{align}
    which is an ($N\!-\!2$)-dimensional affine subspace in $\mathbb{R}^{N}$ and thus is a convex set \cite{Boyd:book2004}. 
    Therefore, $f_{3,\mathrm{rms}}$ preserves convexity on its feasible set, 
    and its iso-surfaces become ($N\!-\!2$)-dimensional ellipsoidal surfaces.
    
    By eliminating the two equality constraints, 
    $f_{3,\mathrm{rms}}$ can be expressed as a convex function defined on $\mathbb{R}^{N-2}$ \cite{Boyd:book2004}:
    \begin{equation}
      f_{3,\mathrm{rms}} = \sqrt{\frac{1}{2}\mathcal{K}^{\mathrm{T}}\mathcal{DK} + \mathcal{G}^{\mathrm{T}}\mathcal{K} + \mathcal{H}},
      \label{eq:f3_rms_new}
    \end{equation}
    where $\mathcal{K}$ is a vector of dimension $N-2$, 
    $\mathcal{D}$ is a positive definite matrix,  
    $\mathcal{G}$ is a constant vector with the same dimension as $\mathcal{K}$, and $\mathcal{H}$ is a constant. 
    For a given lattice, 
    we can fit the parameters $\mathcal{D}$, $\mathcal{G}$, and $\mathcal{H}$, 
    and then analytically obtain the solution with the minimum $f_{3,\mathrm{rms}}$, 
    i.e., the center of ellipsoid. 
    This allows us to generate a Gaussian distribution of $f_{3,\mathrm{rms}}$ in the sextupole strength space, 
    which will be used to develop a fast DA optimization method described in the next section. 
    When the $N$ sextupoles are grouped into $M$ families,
    the analysis above remains applicable in the strength space of sextupole families.

    \begin{figure}[htbp]
      \centering
      \begin{minipage}{\columnwidth}
          \centering
          \includegraphics[width=\columnwidth]{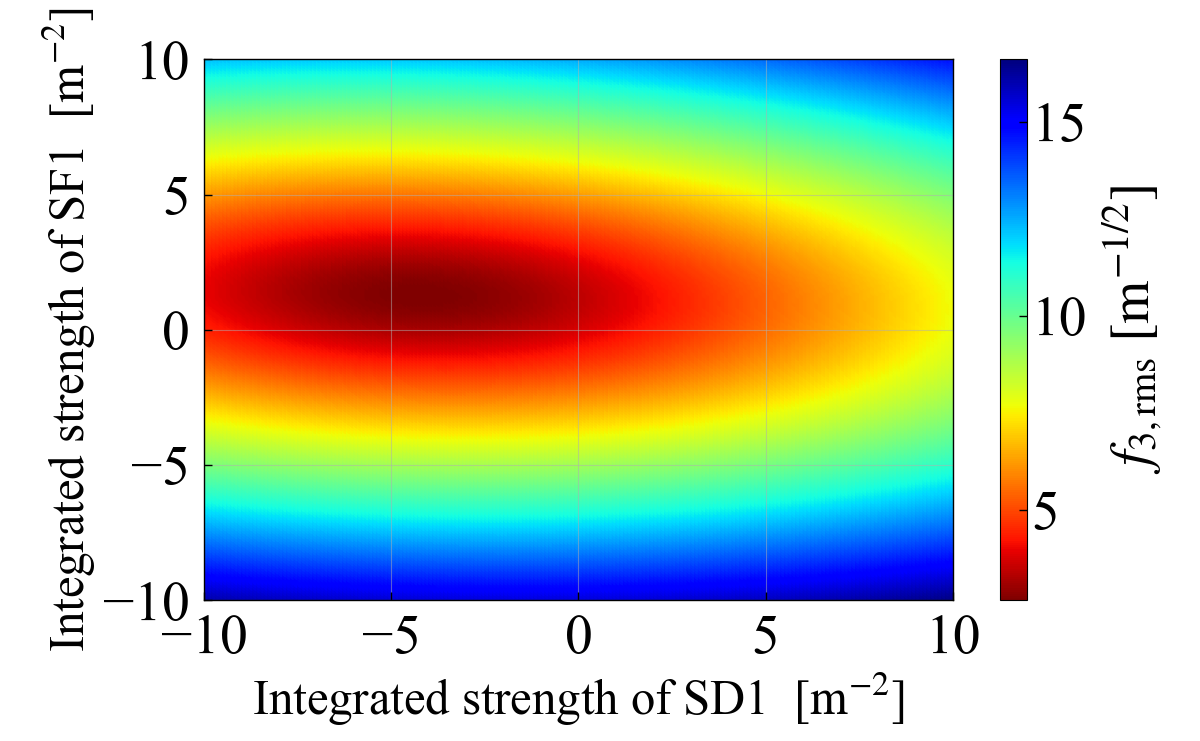}
          \par\medskip
      \end{minipage}
      \begin{minipage}{\columnwidth}
          \centering
          \includegraphics[width=\columnwidth]{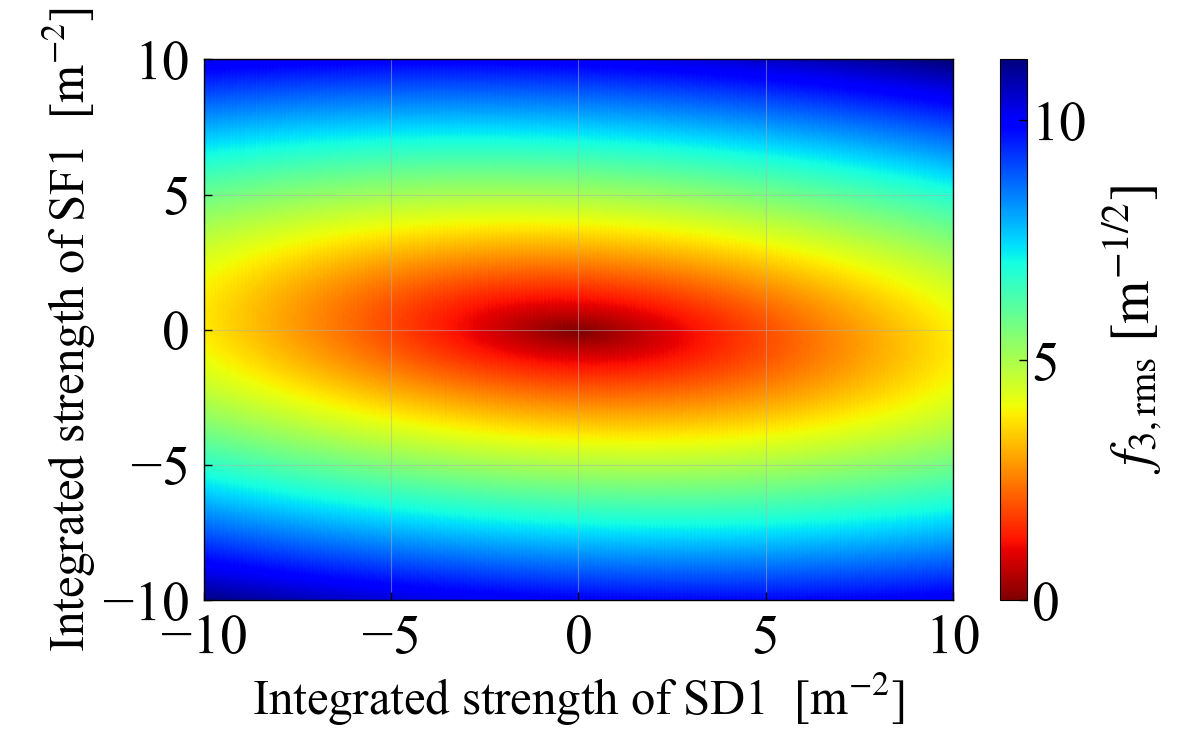}
          \par\medskip
      \end{minipage}
      \caption{
          Distributions of $f_{3,\mathrm{rms}}$ in the normalized integrated strength space of the sextupole families SD1 and SF1 for the FODO lattice with (upper plot) and without (lower plot) the chromaticity correction constraint. 
          }
      \label{fig:FODO_f3_2D}
    \end{figure}

    Taking the FODO lattice as an example, 
    we scanned the normalized integrated strengths of the sextupole families SD1 and SF1 to obtain their $f_{3,\mathrm{rms}}$ values. 
    Figure~\ref{fig:FODO_f3_2D} shows the distributions of $f_{3,\mathrm{rms}}$ in the strength space of SD1 and SF1 for both the cases with and without the chromaticity correction constraint. 
    For the constrained case, the horizontal and vertical chromaticities were corrected to (1.0, 1.0) with the other two sextupole families SD2 and SF2.
    While for the unconstrained case, SD2 and SF2 were not used. 
    It is clearly seen from Fig.~\ref{fig:FODO_f3_2D} that the contours of $f_{3,\mathrm{rms}}$ in both cases are ellipses, 
    in agreement with the theoretical analysis above. 
    Compared to the unconstrained case, 
    the center of the ellipses in the constrained case described by Eq.~\eqref{eq:f3_rms_new} shifts from the origin, 
    and their shapes also change. 

    In addition, $|f_{jklm}(z)|$, $f_{jklm,\mathrm{rms}}$, 
    and $h_{3,\text{ring}}$ have a similar mathematical form to that of $f_{3,\mathrm{rms}}$,  
    which implies that they are also convex functions with ellipsoidal iso-surfaces. 
    However, we found that the longitudinal variation of fourth-order RDTs, 
    quantified in the same form as Eq.~\eqref{eq:f3_rms}, is not a convex function. 

\section{\label{sec:fourth section}Application}
  The convex function $f_{3,\mathrm{rms}}$ plays a crucial role in the non-convex DA optimization. 
  Therefore, the optimization of DA can be regarded as a roughly approximate convex optimization problem, 
  which will be demonstrated through numerical scanning of the HLS-III storage ring lattice. 
  Furthermore, based on the convexity of $f_{3,\mathrm{rms}}$, 
  a fast numerical method for DA optimization will be developed and applied to the SSRF storage ring lattice.

  \subsection{HLS-III lattice} 
    \begin{figure}[htbp]
      \includegraphics[width=0.45\textwidth]{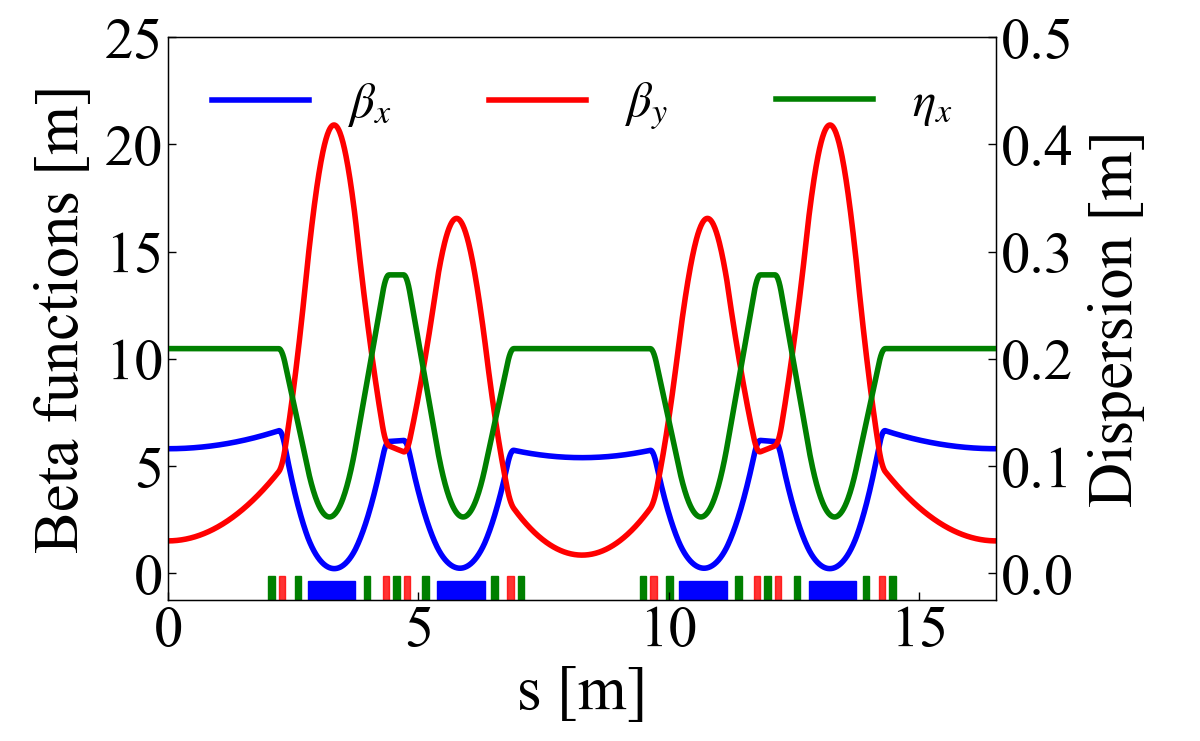}
      \caption{
        Linear optical functions and magnet layout of one lattice period of the HLS-III storage ring. 
        In the magnet layout, bends are represented in blue, quadrupoles in red, and sextupoles in green.
            }
      \label{fig:ddb_lattice}
    \end{figure}

    To compare the distributions of $f_{3,\mathrm{rms}}$ and DA in the sextupole strength space, 
    scanning of sextupole strengths was performed using the HLS-III storage ring lattice \cite{Li:ipac2025}. 
    For each nonlinear solution, $f_{3,\mathrm{rms}}$ was calculated, and DA was tracked through FMA. 
    The total frequency diffusion rate of all surviving particles \cite{Sun:prab2012}, denoted as $\sum d_{r}$, was used to quantitatively represent the DA, 
    as it characterizes not only the size of DA but also the stability of particle motion. 
    Figure~\ref{fig:ddb_lattice} presents the linear optical functions and magnet layout of one lattice cell of the HLS-III storage ring. 
    In order to reduce the computational cost of scanning, the original seven families of sextupoles were regrouped into fewer families. 

\begin{figure}[htbp]
      \centering
      \begin{minipage}{\columnwidth}
          \centering
          \includegraphics[width=0.95\textwidth]{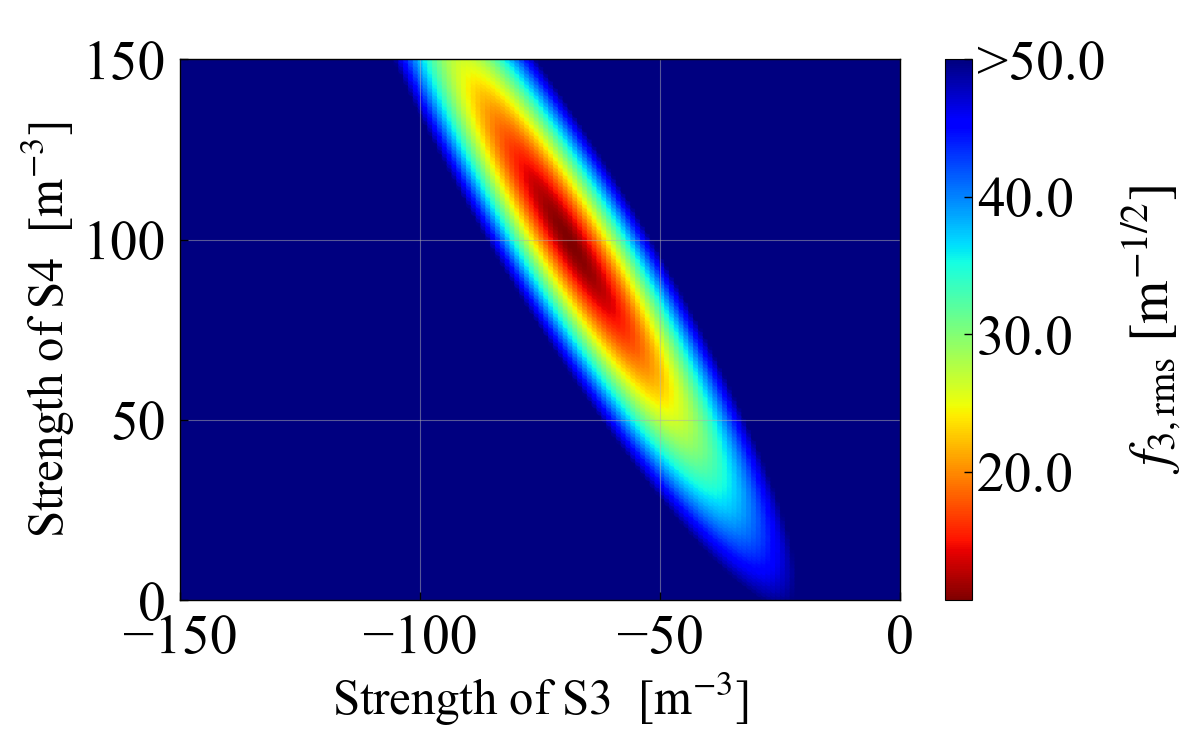}
          \par\medskip
      \end{minipage}
      \begin{minipage}{\columnwidth}
          \centering
          \includegraphics[width=0.95\textwidth]{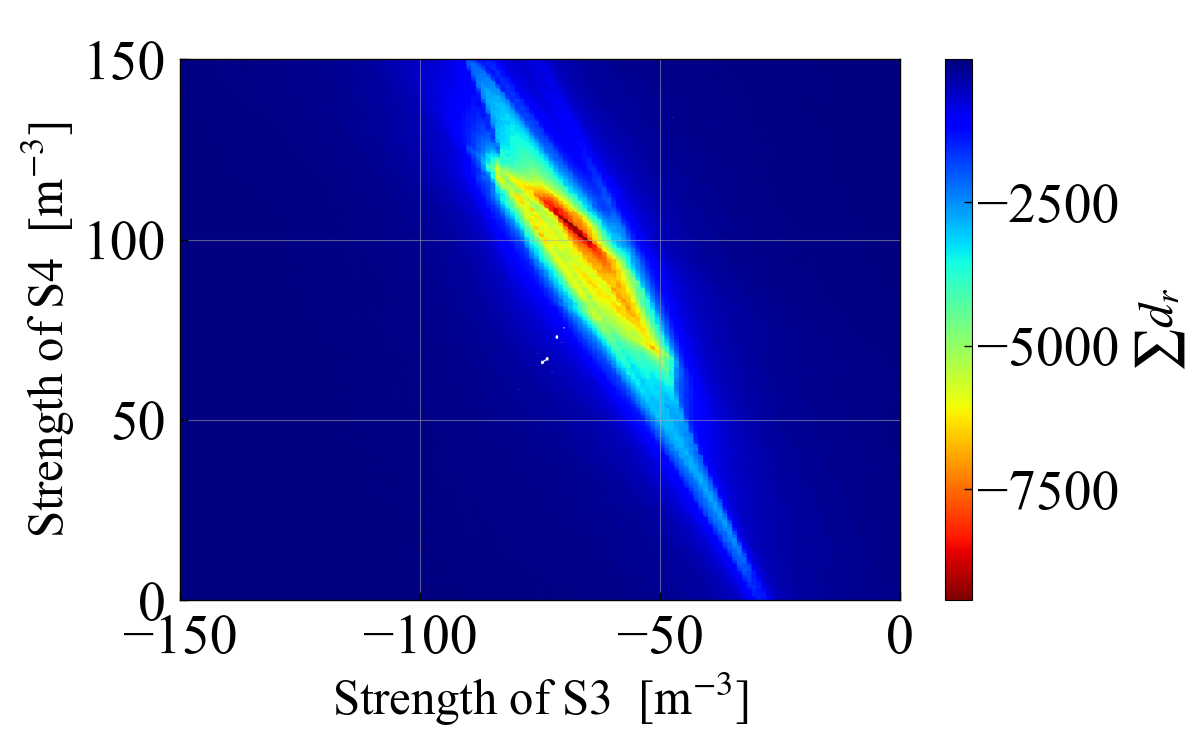}
          \par\medskip
      \end{minipage}
      \caption{
          Distributions of $f_{3,\mathrm{rms}}$ (upper plot) and $\sum d_{r}$ (lower plot) in the normalized strength space of the sextupole families S3 and S4 for the HLS-III lattice with four families of sextupoles. 
          In the upper plot, points with $f_{3,\mathrm{rms}}$ values greater than $50$ are shown in dark blue.
      }
      \label{fig:ddb_f3_Dr}
    \end{figure}

    We firstly considered four families of sextupoles: 
    two (S1, S2) located in the straight sections and two (S3, S4) in the arc sections. 
    In the scanning, the sextupoles S3 and S4 served as free variables, 
    with the horizontal and vertical chromaticities corrected to (1.5, 1.5) with the sextupoles S1 and S2.
    The normalized strengths of S3 and S4 were scanned over the ranges of (-150, 0) $\mathrm{m^{-3}}$ and (0, 150) $\mathrm{m^{-3}}$, respectively.
    Figure~\ref{fig:ddb_f3_Dr} shows the distributions of $f_{3,\mathrm{rms}}$ and $\sum d_{r}$ in the strength space of the sextupoles S3 and S4. 
    It is seen that the contours of $f_{3,\mathrm{rms}}$ are a series of concentric and coaxial ellipses,  
    and that the distributions of $f_{3,\mathrm{rms}}$ and $\sum d_{r}$ exhibit strong consistency. 
    Specifically, regions with smaller $f_{3,\mathrm{rms}}$ values broadly coincide with those with lower $\sum d_{r}$, 
    and their variation trends are roughly the same. 

\begin{figure*}[htbp]
      \centering
      \begin{minipage}[b]{0.475\textwidth}
          \centering
          \includegraphics[width=\linewidth]{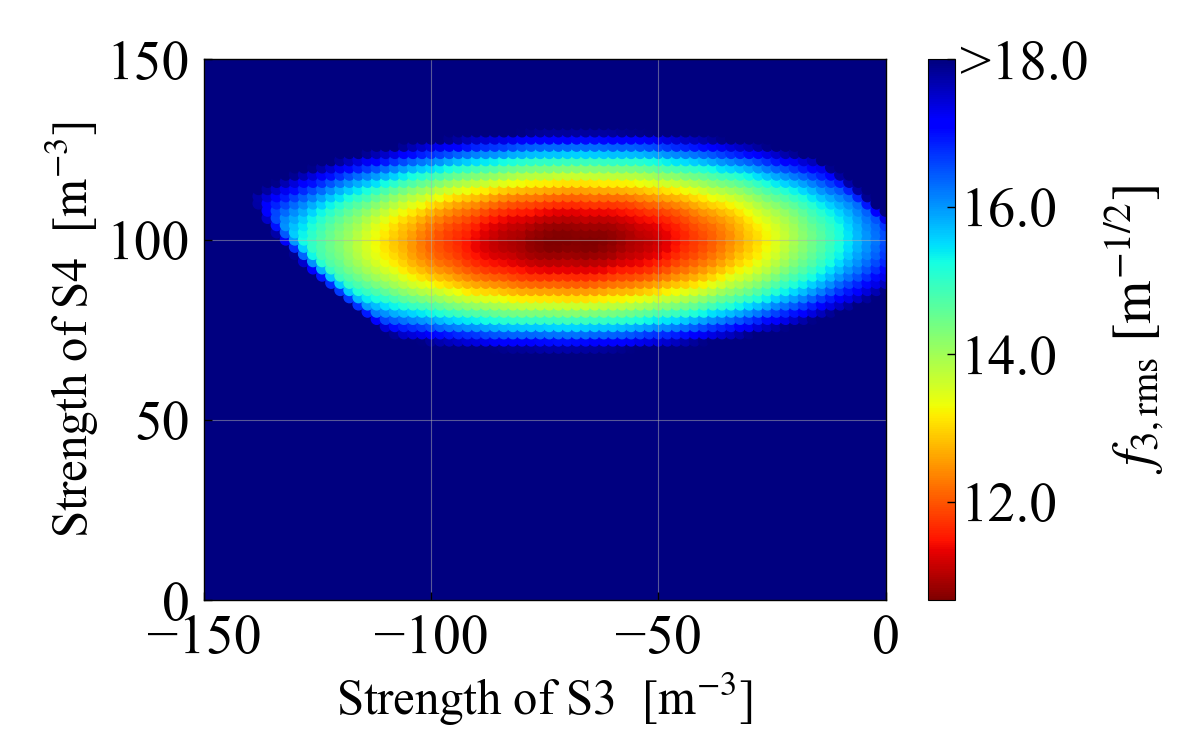}
          \par\medskip
      \end{minipage}
      \hspace{0.25em}
      \begin{minipage}[b]{0.475\textwidth}
          \centering
          \includegraphics[width=\linewidth]{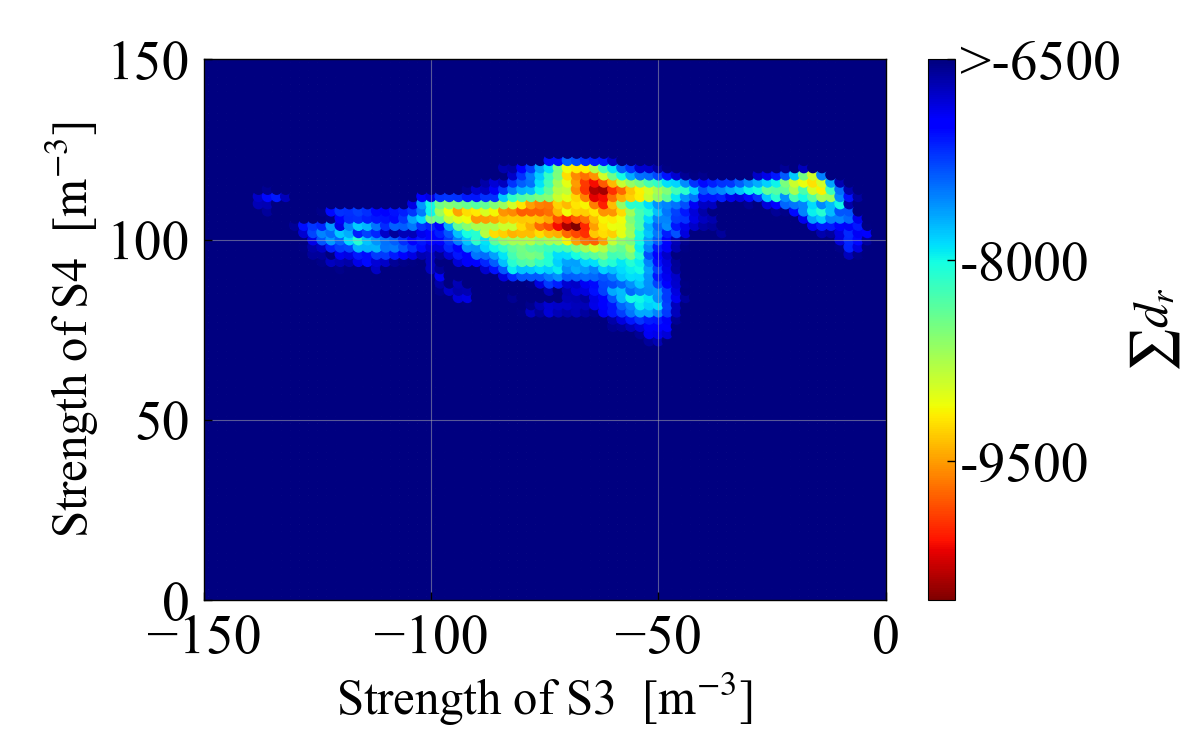}
          \par\medskip
      \end{minipage}
      \begin{minipage}[b]{0.475\textwidth}
          \centering
          \includegraphics[width=\linewidth]{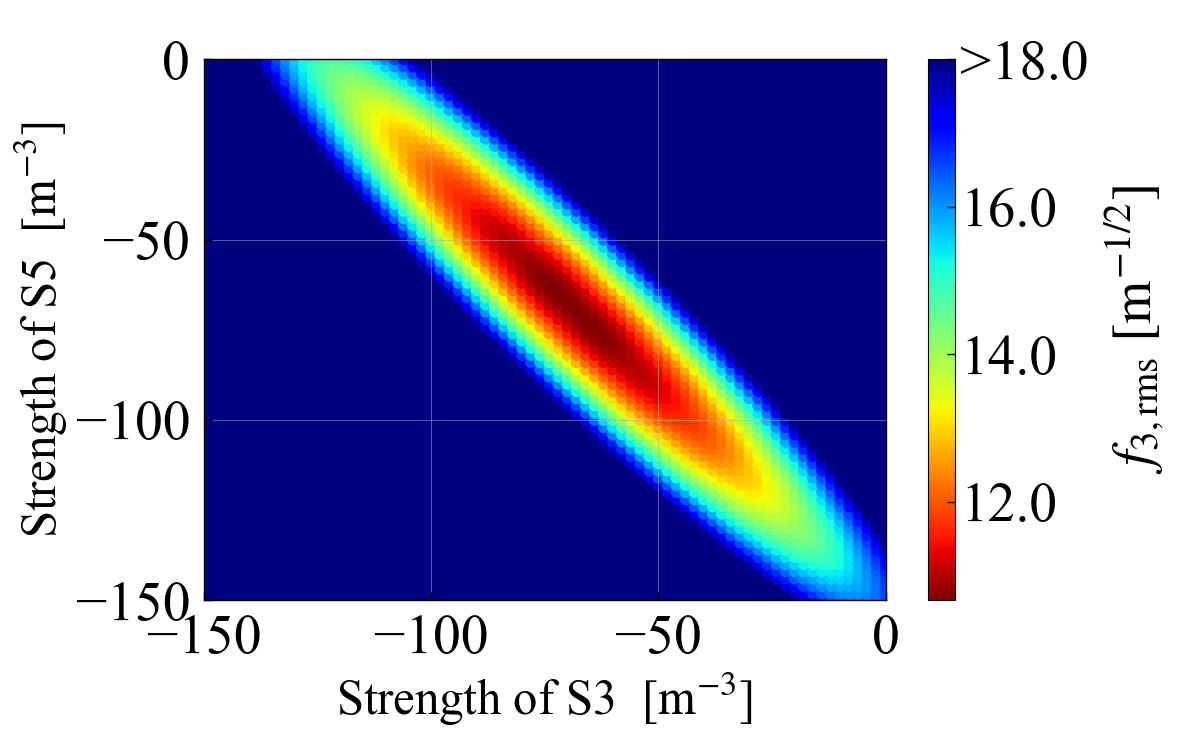}
          \par\medskip
      \end{minipage}
      \hspace{0.25em}
      \begin{minipage}[b]{0.475\textwidth}
          \centering
          \includegraphics[width=\linewidth]{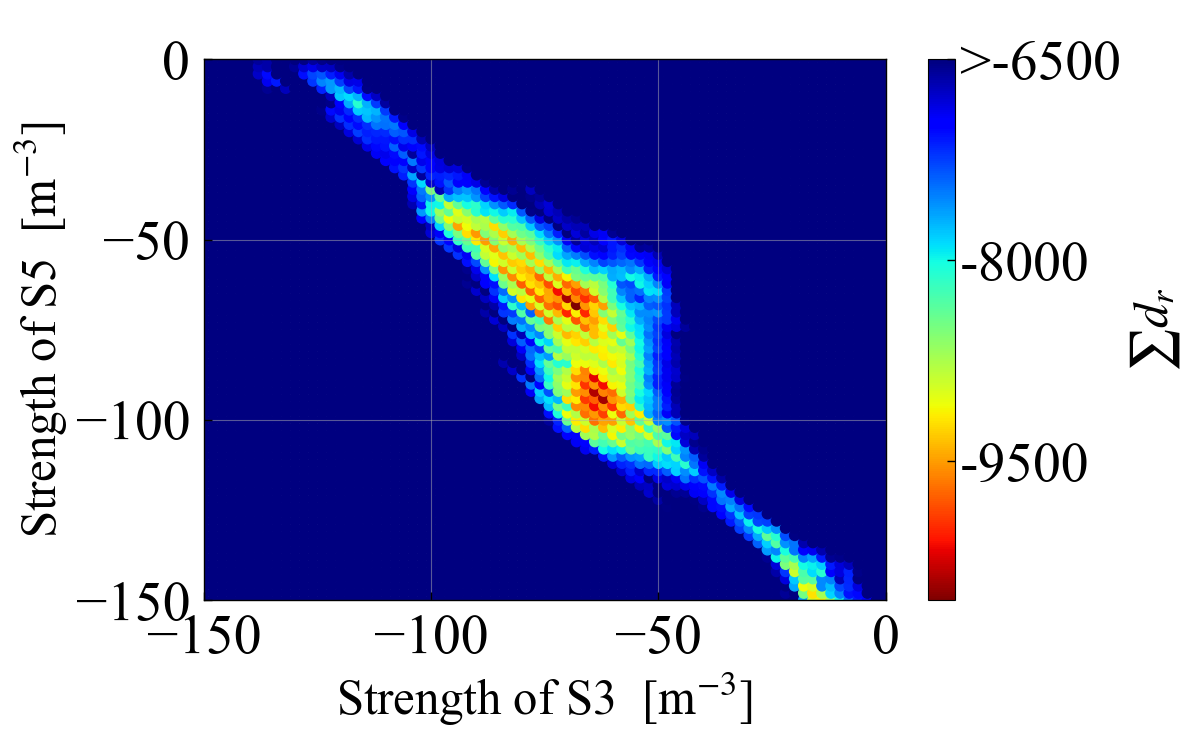}
          \par\medskip
      \end{minipage}
      \begin{minipage}[b]{0.475\textwidth}
          \centering
          \includegraphics[width=\linewidth]{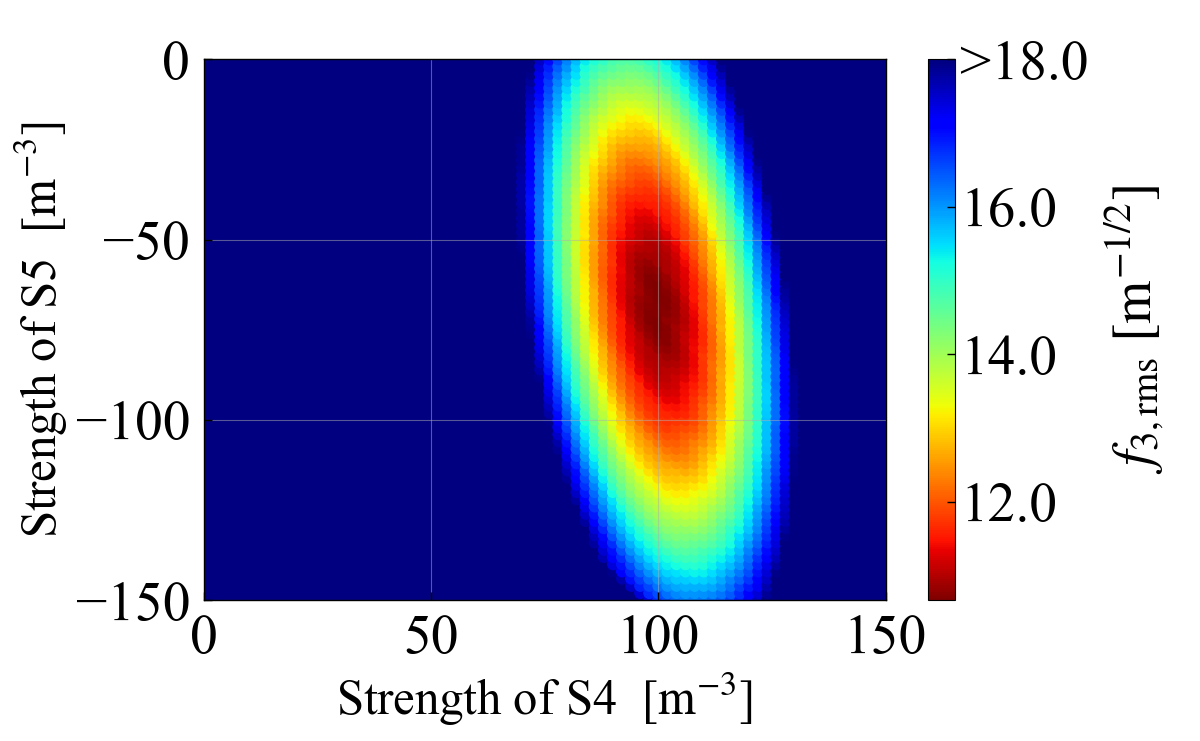}
          \par\medskip
      \end{minipage}
      \hspace{0.25em}
      \begin{minipage}[b]{0.475\textwidth}
          \centering
          \includegraphics[width=\linewidth]{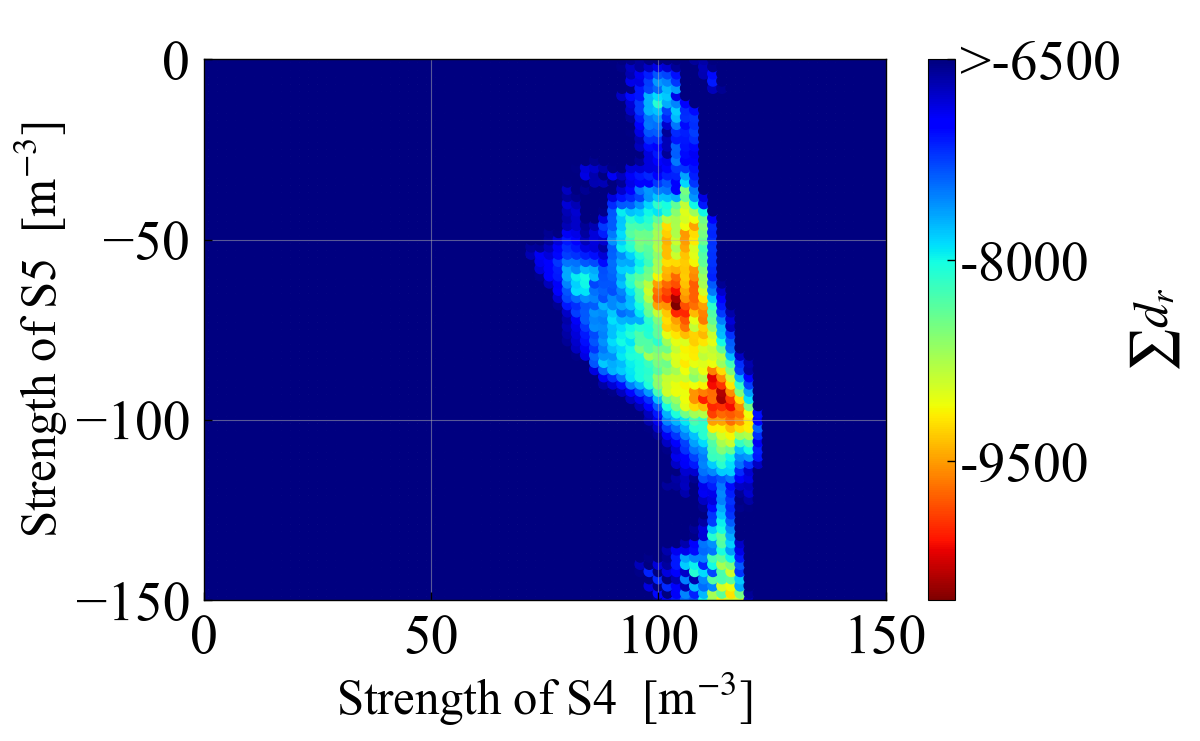}
          \par\medskip
      \end{minipage}
      
      \caption{      
        Projected distributions of $f_{3,\mathrm{rms}}$ and $\sum d_{r}$ in the normalized strength spaces of the sextupole families S3 and S4 (two top plots), 
        S3 and S5 (two middle plots), and S4 and S5 (two bottom plots) for the HLS-III lattice with five families of sextupoles. 
        The projection from the three-variable space ($\mathrm{S}_{i}$, $\mathrm{S}_{j}$, $\mathrm{S}_{k}$) onto the two-variable space ($\mathrm{S}_{i}$, $\mathrm{S}_{j}$) is formed by taking, for each ($\mathrm{S}_{i}$, $\mathrm{S}_{j}$), 
        the point that has the smallest $f_{3,\mathrm{rms}}$ or lowest $\sum d_{r}$ value in the $\mathrm{S}_{k}$ dimension. 
        The incompleteness of the outer ellipses of $f_{3,\mathrm{rms}}$, 
        as shown in the top left plot, is due to the limitation of the variable ranges. 
        In the three left plots, points with $f_{3,\mathrm{rms}}$ values greater than 18 are shown in dark blue; 
        in the three right plots, points with $\sum d_{r}$ values higher than -6500 are shown in dark blue.
        }
      \label{fig:DDB_f3_Dr}
    \end{figure*}

    Then, we studied the case with five sextupole families: 
    three (S3, S4, S5) in the arc sections as free variables, 
    and two (S1, S2) in the straight sections for correcting the chromaticities to (1.5, 1.5). 
    The normalized strengths were scanned over the ranges of (-150, 0) $\mathrm{m^{-3}}$ for S3 and S5, 
    and (0, 150) $\mathrm{m^{-3}}$ for S4. 
    Figure~\ref{fig:DDB_f3_Dr} shows the projected distributions of $f_{3,\mathrm{rms}}$ and $\sum d_{r}$ in the strength spaces of S3 and S4, S3 and S5, and S4 and S5. 
    It is seen that the consistency between the distributions of $f_{3,\mathrm{rms}}$ and $\sum d_{r}$ remains strong. 
    Although $\sum d_{r}$ has two extreme regions (i.e., the red regions in the figure), 
    they are not discrete ``islands” but distributed on the same ``continent”, 
    which is located within the central region of the ellipsoid of $f_{3,\mathrm{rms}}$. 
    This differs from the common understanding that DA may have a distribution of multiple extreme ``islands". 

\begin{figure*}[htbp]
      \centering
      \begin{minipage}[b]{0.475\textwidth}
          \centering
          \includegraphics[width=\linewidth]{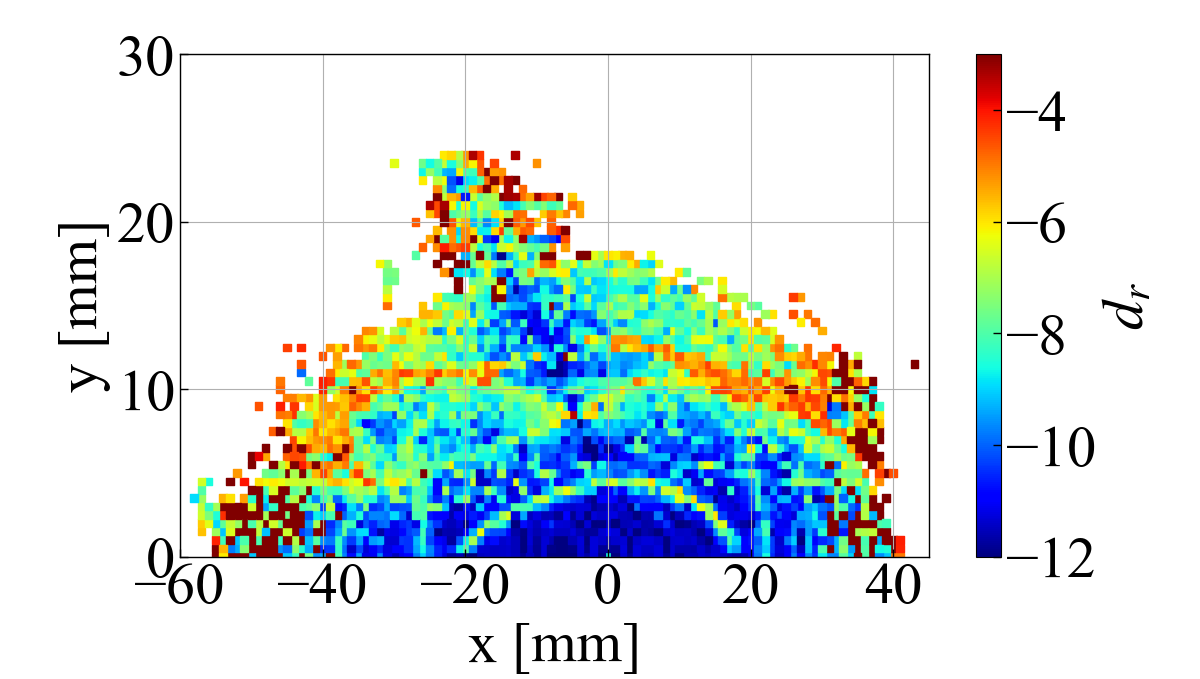}
          \par\medskip
      \end{minipage}
      \hspace{0.15em}
      \begin{minipage}[b]{0.475\textwidth}
          \centering
          \includegraphics[width=\linewidth]{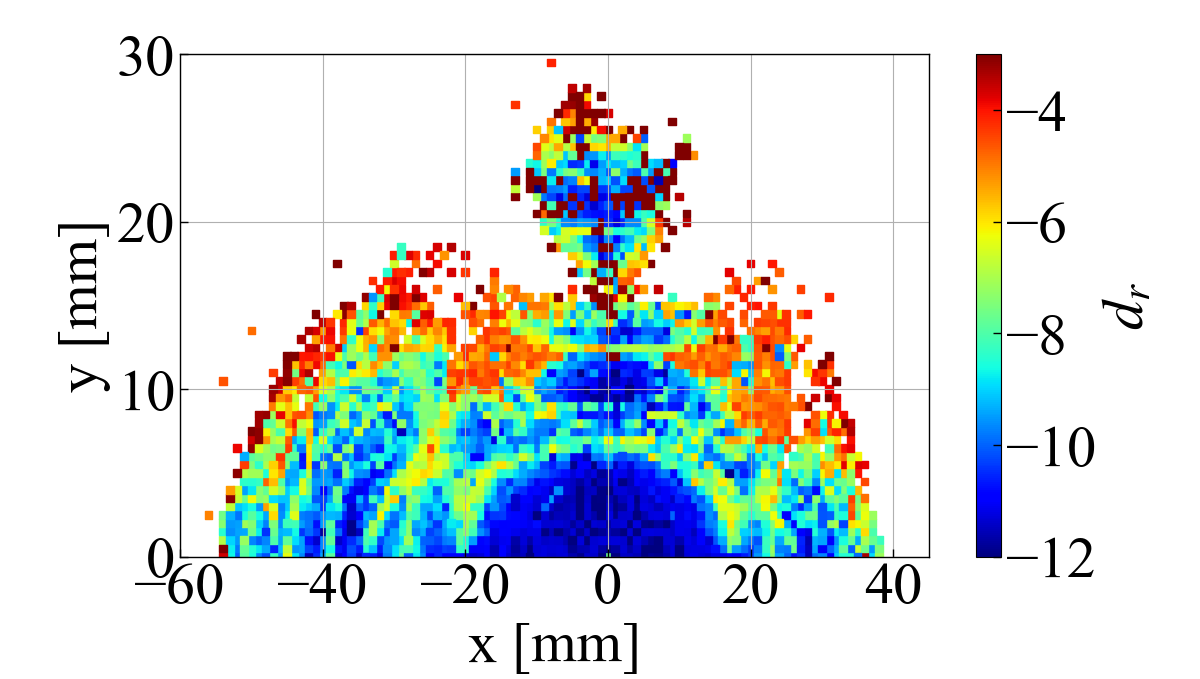}
          \par\medskip
      \end{minipage}
      \begin{minipage}[b]{0.475\textwidth}
          \centering
          \includegraphics[width=\linewidth]{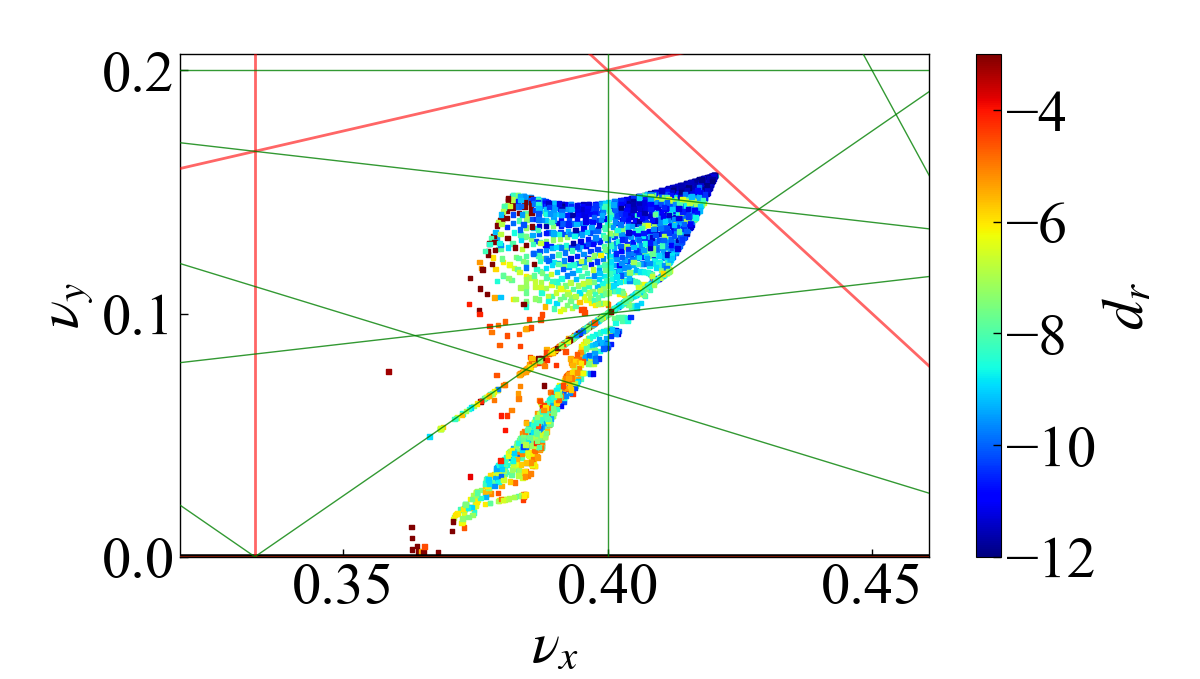}
          \par\medskip
      \end{minipage}
      \hspace{0.15em}
      \begin{minipage}[b]{0.475\textwidth}
          \centering
          \includegraphics[width=\linewidth]{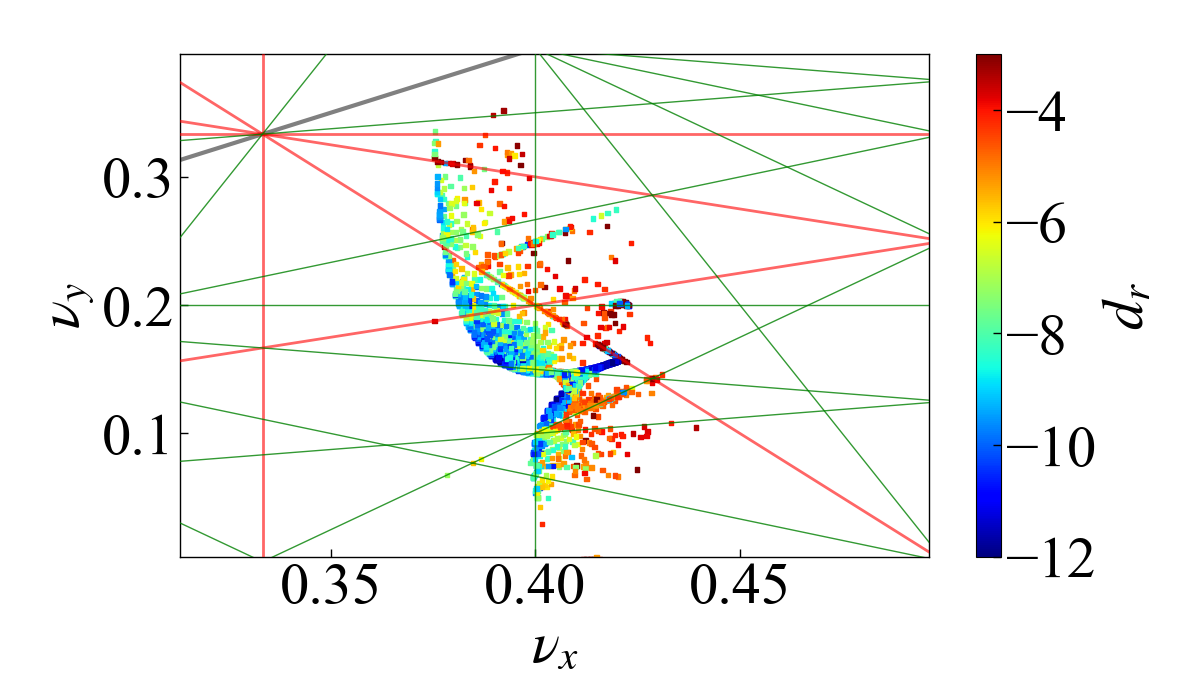}
          \par\medskip
      \end{minipage}
      \begin{minipage}[b]{0.475\textwidth}
          \centering
          \includegraphics[width=\linewidth]{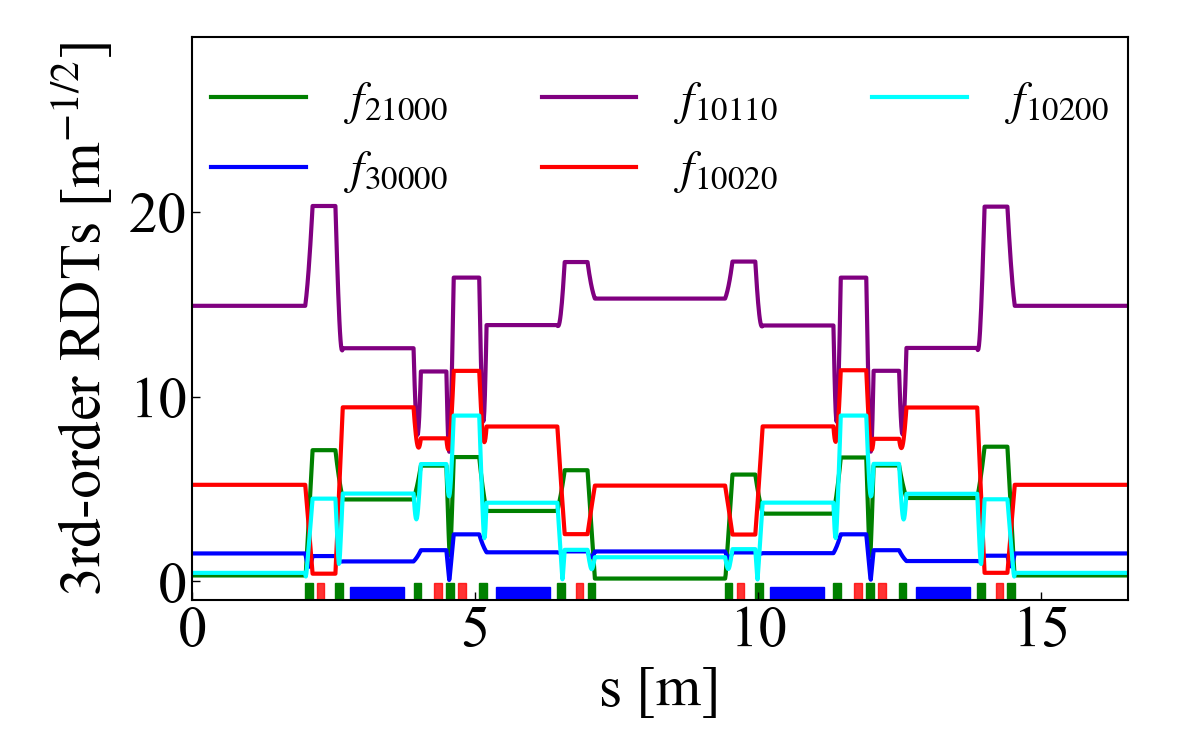}
          \par\medskip
      \end{minipage}
      \hfill
      \begin{minipage}[b]{0.475\textwidth}
          \centering
          \includegraphics[width=\linewidth]{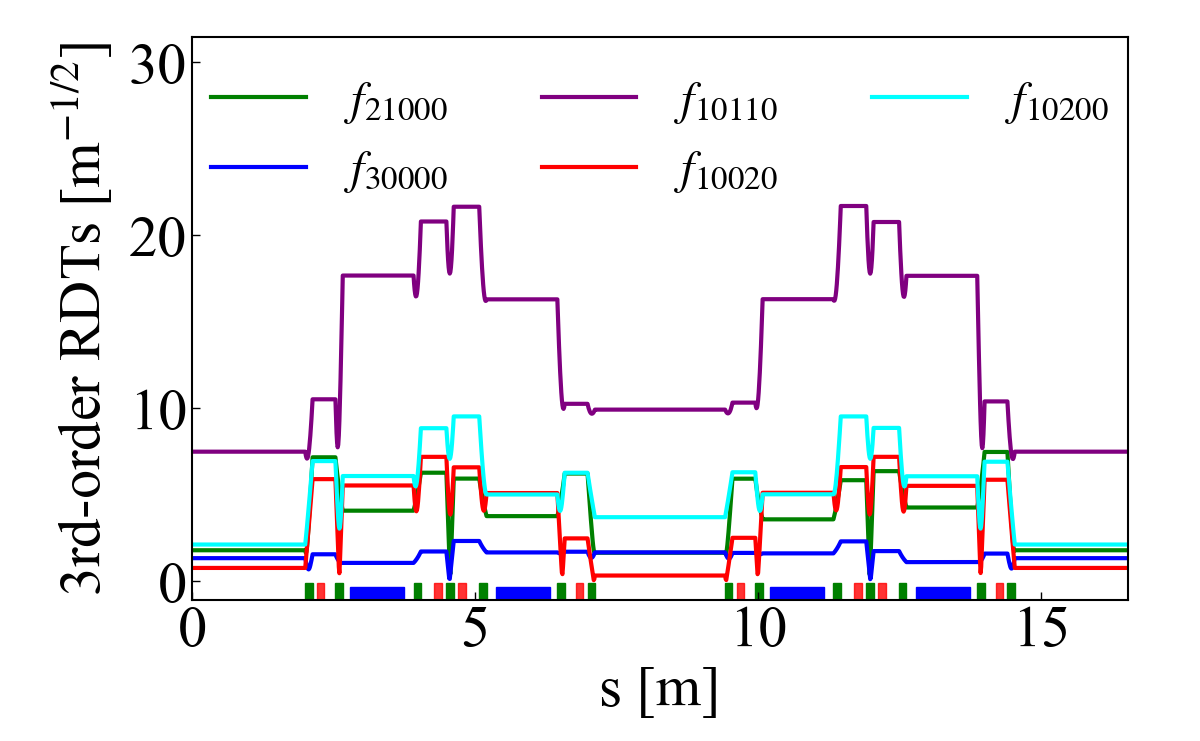}
          \par\medskip
      \end{minipage}
      
      \caption{
          FMA of DAs (four upper plots) and longitudinal variations of third-order RDTs (two lower plots) for the two selected nonlinear solutions from Fig.~\ref{fig:DDB_f3_Dr}. 
          The three left plots are for Solution 1 in Table~\ref{tab:parameters}, and the right for Solution 2. 
      }
      \label{fig:two_nonlinear_solutions}
    \end{figure*}

    Two nonlinear solutions from the two extreme regions were selected, 
    with their sextupole strengths, $f_{3,\mathrm{rms}}$, and $\sum d_{r}$ listed in Table~\ref{tab:parameters}. 
    Figure~\ref{fig:two_nonlinear_solutions} presents the FMA of their DAs and the longitudinal variations of their third-order RDTs.
    While the sextupole strengths differ between the two solutions, particularly for S2 and S5, 
    they have comparable $f_{3,\mathrm{rms}}$ values and DA performance.  
    Recently, a chaos suppression method was applied to optimize the DA of the NSLS-II storage ring lattice, 
    yielding a solution with significantly different sextupole strengths but a similar DA compared to the reference solution \cite{Liyongjun:prab2026arxiv}. 
    The longitudinal variations of third-order RDTs for both solutions are also comparable \cite{Liyongjun:prab2026arxiv}. 
    These results indicate that distinct good solutions with comparable DAs are typically connected through the ellipsoidal surface of $f_{3,\mathrm{rms}}$.

    \begin{table}[htbp]
      \centering
      \caption{\label{tab:parameters}
      The normalized strengths of sextupoles, $f_{3,\mathrm{rms}}$, and $\sum d_{r}$ for the two selected nonlinear solutions from Fig.~\ref{fig:DDB_f3_Dr}. }
      \begin{ruledtabular}
        \begin{tabular}{crrc}
          Parameter & Solution 1 & Solution 2 & Unit\\
          \colrule
          Strength of S1 & 53.60 & 56.84 & $\mathrm{m}^{-3}$\\
          Strength of S2 & -60.47 & -76.60 & $\mathrm{m}^{-3}$\\
          Strength of S3 & -58.00 & -64.00 & $\mathrm{m}^{-3}$\\
          Strength of S4 & 102.00 & 98.00 & $\mathrm{m}^{-3}$\\
          Strength of S5 & -92.00 & -62.00 & $\mathrm{m}^{-3}$\\
          $f_{3,\mathrm{rms}}$ & 16.52 & 16.66 & $\mathrm{m}^{-1/2}$\\
          $\sum d_{r}$ & -8919.80 & -8826.80 & -- \\
        \end{tabular}
      \end{ruledtabular}
    \end{table}

    Combining the scan results from the two cases with different numbers of sextupole families,
    it is observed that the distributions of $f_{3,\mathrm{rms}}$ and $\sum d_{r}$ are largely consistent, 
    and that the extreme regions of $\sum d_{r}$ are enclosed by the ellipsoidal contour with smaller $f_{3,\mathrm{rms}}$. 
    Since $f_{3,\mathrm{rms}}$ is a convex function and has a very strong relationship with DA, 
    it is reasonable to infer that in a higher-dimensional variable space, 
    the distributions of $f_{3,\mathrm{rms}}$ and DA will also exhibit good consistency. 
    That is, better DA solutions are also distributed within the central region of the $f_{3,\mathrm{rms}}$ ellipsoid. 
    Therefore, this suggests that, based on the distribution of $f_{3,\mathrm{rms}}$, 
    more solutions can be generated in the ellipsoid with smaller $f_{3,\mathrm{rms}}$ values to serve as the initial population of an intelligent algorithm for improving the performance of particle tracking based DA optimization.

  \subsection{SSRF lattice}
    In the numerical optimization of DA using intelligent algorithms, 
    the initial population is generally randomly distributed within the variable space. 
    Based on the roughly approximate convexity of DA and the ellipsoidal geometry of $f_{3,\mathrm{rms}}$, 
    a high-quality initial population can be generated to improve optimization performance. 
    It is observed from Eq.~\eqref{eq:f3_rms_new} that $f_{3,\mathrm{rms}}^{2}$ is a standard quadratic function, 
    which resembles the exponent part of the multivariate Gaussian probability density function. 
    This leads us to initialize the population with a Gaussian distribution that is geometrically aligned with the ellipsoid of $f_{3,\mathrm{rms}}^{2}$. 
    The population initialization steps are as follows: 
    (1) Determine the parameters $\mathcal{D}$, $\mathcal{G}$, and $\mathcal{H}$ in Eq.~\eqref{eq:f3_rms_new} for $f_{3,\mathrm{rms}}^{2}$ by fitting with some randomly generated nonlinear solutions, 
    and then obtain the central position with minimum $f_{3,\mathrm{rms}}^{2}$; 
    (2) Utilize the central position and $\mathcal{D}$ to construct the probability density function for a Gaussian distribution, with $3\sigma$ equal to $\lambda$ times minimum $f_{3,\mathrm{rms}}^{2}$, where $\lambda\!>\!1$; 
    (3) Initialize the population using this Gaussian distribution in the variable space.
    
\begin{figure}[htbp]
      \includegraphics[width=0.45\textwidth]{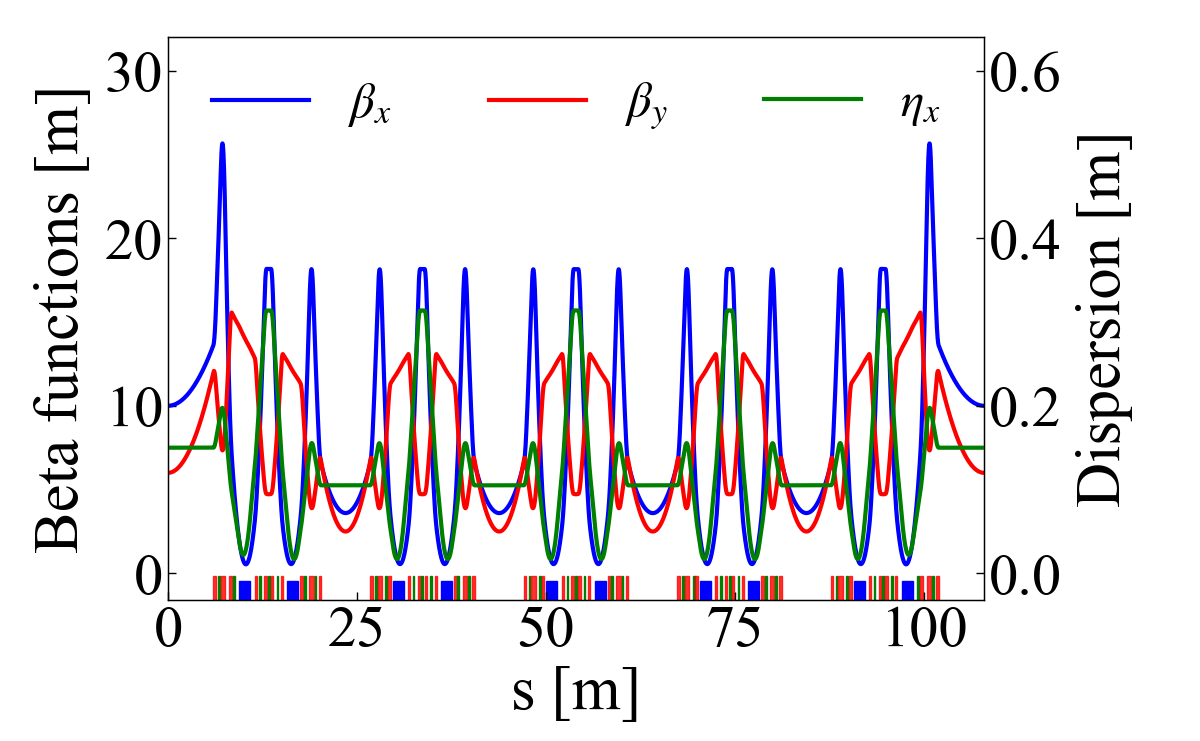}
      \caption{Linear optical functions and magnet layout of one lattice super-period of the SSRF storage ring.}
      \label{fig:ssrf_lattice}
    \end{figure}

    The SSRF storage ring lattice with eight families of sextupoles \cite{Liu:cpc2006} was used to conduct the DA optimization with Gaussian-distributed initial population. 
    The lattice of one super-period of the SSRF storage ring is presented in Fig.~\ref{fig:ssrf_lattice}. 
    A differential evolution algorithm \cite{Husain:nima2018} with a population size of 50 was iterated for 50 generations to perform the optimization. 
    The factor $\lambda$ was set to 1.25 in the generation of Gaussian-distributed initial population. 
    The optimization variables are the strengths of the six sextupole families in the straight sections, 
    with the two families of sextupoles in the arc sections correcting the horizontal and vertical chromaticities to (1.0, 1.0). 
    The on-momentum DA was optimized by minimizing $\sum d_{r}$ as the single optimization objective. 
    As a comparison, the DA optimization with randomly distributed initial population was also conducted, 
    with the same algorithm settings, optimization variables, and optimization objective. 
    To reduce the impact of the randomness of algorithm, 
    each of the two optimizations was performed three times independently. 

\begin{figure}[htbp]
      \centering
      \begin{minipage}{\columnwidth}
          \centering
          \includegraphics[width=0.95\textwidth]{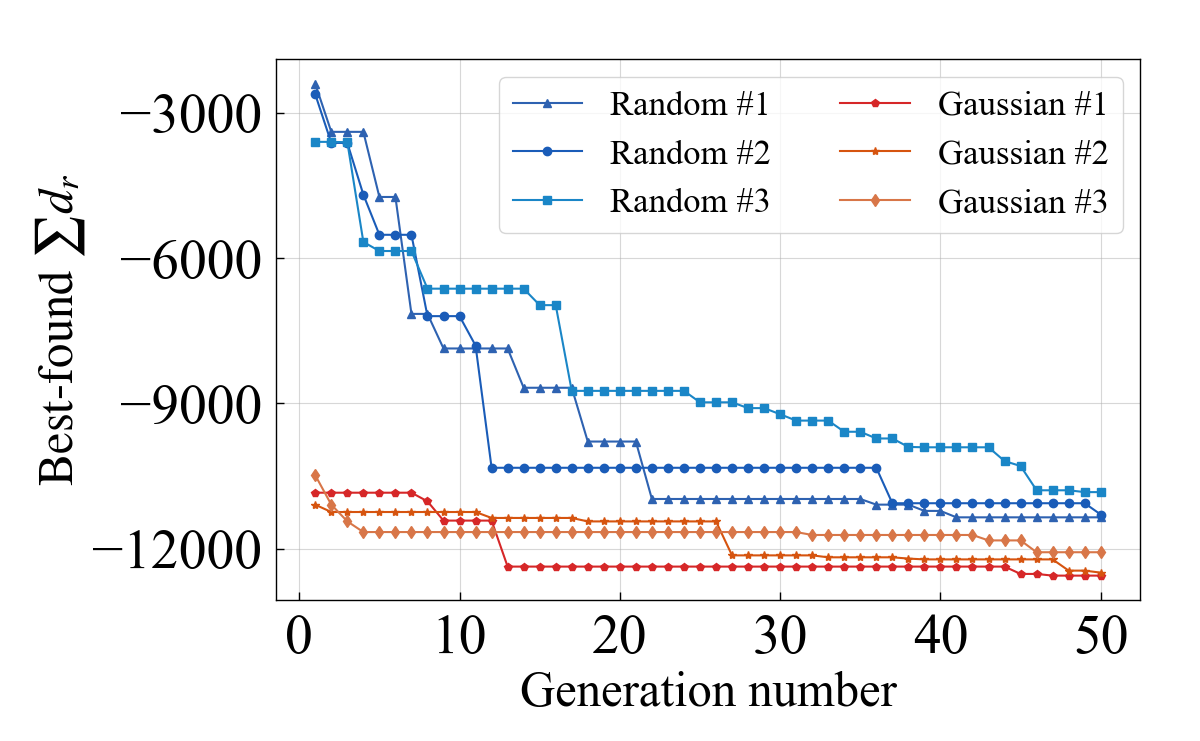}
          \par\medskip
      \end{minipage}
      \begin{minipage}{\columnwidth}
          \centering
          \includegraphics[width=0.95\textwidth]{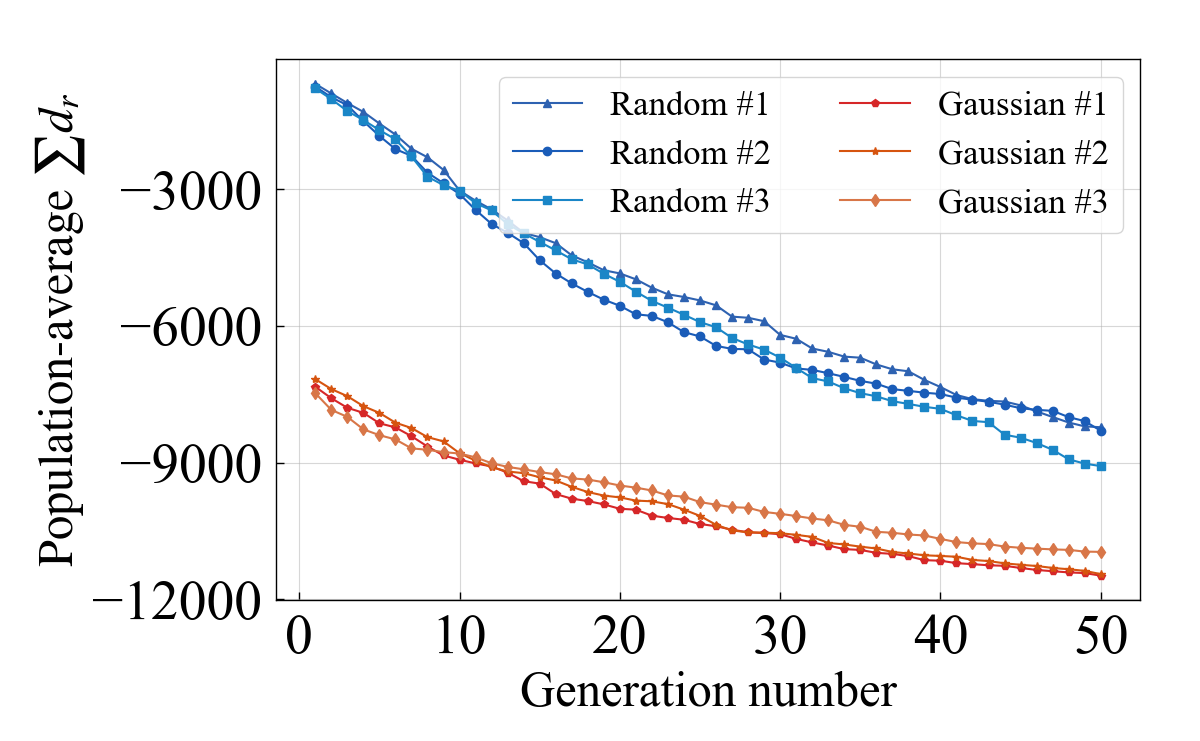}
          \par\medskip
      \end{minipage}
      \caption{
          Evolutions of the best-found $\sum d_{r}$ (upper plot) and the population-average $\sum d_{r}$ (lower plot) with the generation number, 
          for Gaussian-distributed and randomly distributed initial populations. 
          Three optimization runs were performed for each kind of distribution. 
      }
      \label{fig:DE_Dr}
    \end{figure}

    Figure~\ref{fig:DE_Dr} presents the convergence curves of the best-found $\sum d_{r}$ and the population-average $\sum d_{r}$ for Gaussian-distributed and randomly distributed initial populations. 
    The results show that the optimization with a Gaussian-distributed initial population converges significantly faster than that with a randomly distributed initial population. 
    In the first generation, the Gaussian case shows significantly better best and average results than the random case. Moreover, its results at the 10th generation even outperform those of the random case at the 50th generation. 
    This clearly demonstrates that adopting a Gaussian distribution substantially improves the quality of initial population, thereby enhancing search efficiency. 
    Furthermore, the optimization results also reflect the strong consistency between the distributions of $f_{3,\mathrm{rms}}$ and DA in a higher-dimensional variable space. 

\section{\label{sec:fifth section}Conclusion and outlook}
    The optimization of DA is crucial for the design of low-emittance storage rings. 
    However, its inherent non‑convex nature makes global optimization very challenging. 
    This paper further investigated the properties of the longitudinal variation of third‑order RDTs, 
    which plays a key role in DA optimization. 
    We proved that the longitudinal variation of third‑order RDTs, 
    quantified by their RMS value $f_{3,\mathrm{rms}}$ at sextupole locations, 
    is a special convex function. 
    The iso-surfaces of $f_{3,\mathrm{rms}}$ in the sextupole strength space are a series of concentric and coaxial ellipsoidal surfaces. 
    This convexity connects the complex non-convex DA optimization problem with a convex function featuring a well-defined geometric structure, 
    providing a new foundation for understanding and accelerating DA optimization. 

    The results from scanning the sextupole strength space of the HLS‑III storage ring lattice showed a strong consistency between the distributions of $f_{3,\mathrm{rms}}$ and DA. 
    Solutions with better DAs are distributed within the central region of the ellipsoid of $f_{3,\mathrm{rms}}$, 
    which is different from the common understanding that DA may have multiple isolated extreme regions. 
    Therefore, DA optimization can be regarded as a roughly approximate convex optimization problem. 
    Based on this, we developed a fast DA optimization method based on particle tracking. 
    In this method, a high-quality initial population for an intelligent algorithm is generated with a Gaussian distribution that follows the geometry of the $f_{3,\mathrm{rms}}^{2}$ ellipsoid. 
    This significantly accelerates convergence, as demonstrated in the SSRF storage ring lattice, 
    compared to the commonly used method with a randomly distributed initial population. 
    In the future, the work will be extended to the optimization of both on- and off-momentum DAs, 
    and we will further explore the simultaneous optimization of linear optics and nonlinear dynamics based on this work. 
   
\begin{acknowledgments}
  This work was supported by the National Natural Science Foundation of China under Grant No. 12375325.
\end{acknowledgments}

\bibliography{PRAB}

\end{document}